\documentclass[pre,tighten]{revtex4}

\usepackage{amssymb,amsmath}

\usepackage{epsfig}

\usepackage{graphicx}% Include figure files
\usepackage{color}

\newcommand\beq{\begin{equation}}
\newcommand\eeq{\end{equation}}
\newcommand\beqa{\begin{eqnarray}}
\newcommand\eeqa{\end{eqnarray}}
\newcommand{\dd}{\text{d}}

\newcommand{\al}{\alpha}

\begin{document}
\title{Tracer diffusion coefficients in a sheared inelastic Maxwell gas}
\author{Vicente Garz\'{o}\footnote[1]{Electronic address: vicenteg@unex.es;
URL: http://www.unex.es/eweb/fisteor/vicente/}}
\affiliation{Departamento de F\'isica and Instituto de Computaci\'on Cient\'{\i}fica Avanzada (ICCAEx), Universidad de Extremadura, E-06071 Badajoz, Spain}
\author{Emmanuel Trizac\footnote[2]{Electronic address: trizac@lptms.u-psud.fr;
URL: http://www.lptms.u-psud.fr/membres/trizac/}} \affiliation{Laboratoire de Physique
Th\'eorique et Mod\`eles Statistiques (CNRS UMR 8626), B$\hat{a}$timent 100,
Universit\'e Paris-Sud, 91405 Orsay cedex, France}

\begin{abstract}
We study the transport properties of an impurity in a sheared granular gas, in the framework of the Boltzmann equation for inelastic Maxwell models.
We investigate here the impact of a nonequilibrium phase transition found in such systems, where the tracer species carries a finite
fraction of the total kinetic energy (ordered phase). To this end, the diffusion coefficients are first obtained for a granular binary mixture in spatially inhomogeneous
states close to the simple shear flow. In this situation, the set of coupled Boltzmann equations are solved by means of a Chapman-Enskog-like expansion around the (local)
shear flow distributions for each species, thereby retaining all the hydrodynamic orders in the shear rate $a$. Due to the anisotropy induced by the shear flow, three tensorial
quantities $D_{ij}$, $D_{p,ij}$, and $D_{T,ij}$ are required to describe the mass transport process instead of the conventional scalar coefficients. These tensors are given in terms of the solutions of a set of
coupled algebraic equations, which can be \emph{exactly} solved as functions of the
shear rate $a$, the coefficients of restitution $\alpha_{sr}$ and the parameters of the mixture (masses and composition). Once the forms of $D_{ij}$, $D_{p,ij}$, and $D_{T,ij}$
are obtained for arbitrary mole fraction $x_1=n_1/(n_1+n_2)$ (where $n_r$ is the number density of species $r$), the tracer limit ($x_1\to 0$) is carefully considered for the
above three diffusion tensors. Explicit forms for these coefficients are derived showing that their shear rate dependence is significantly affected by the order-disorder transition.
\end{abstract}

\date{\today}
\maketitle

\section{Introduction}
\label{sec1}

The model of smooth inelastic hard spheres (IHS) has proven insightful to characterize the influence of collisional dissipation on the dynamic properties of rapid granular
flows \cite{BP04,P14}. Following this minimal route, the inelasticity of collisions is accounted for by a constant (positive) coefficient of normal restitution $\al \leq 1$
that only impinges on the translational degrees of freedom of grains \cite{BP04}. The case $\al=1$ stems for elastic, kinetic energy conserving collisions. On the other hand, the complex mathematical structure of the Boltzmann collision operator for IHS prevents us from obtaining exact results, even in the simplest homogeneous cooling state. To get the explicit forms of the Navier-Stokes transport coefficients \cite{NS} one usually considers
the leading order in a Sonine polynomial
expansion of the velocity distribution function \cite{BP04,P14}. These difficulties increase considerably when one studies
multicomponent systems (namely, a mixture of grains with different masses, sizes
and coefficients of restitution) since not only the number of transport coefficients is larger than for a single gas but also the kinetic description involves a set of coupled Boltzmann
equations for the one-particle velocity distribution function of each species.

One of the main mathematical intricacies in evaluating the collisional moments of the Boltzmann operator for hard spheres (even for ordinary mixtures) comes from the fact that the collision
rate is proportional to the magnitude of the relative velocity of the two colliding spheres. This property precludes the possibility of determining those collisional moments without
the knowledge of the velocity distribution functions. In the case of elastic fluids, a possible
way to overcome this problem (keeping the structure of the Boltzmann collision operator) is to assume that the particles interact via
the repulsive Maxwell potential (inversely proportional to the fourth power
of the distance). For this interaction model, the collision rate is independent
of the relative velocity and this brings a number of convenient mathematical
properties of the Boltzmann collision operator \cite{E81}. Thanks to this
simplification, nonlinear transport properties can be exactly obtained \cite{TM80,GS03} from the Boltzmann equation for Maxwell elastic molecules and, when
properly reduced, they exhibit a good agreement with results obtained for
other interaction models \cite{Hess}. In the context of inelastic gases, the Boltzmann
equation for inelastic Maxwell models (IMM) was also introduced about
sixteen years ago \cite{Maxwell}. The IMM share with elastic Maxwell molecules the
property that the collision rate is velocity \emph{independent} but their collision
rules are the same as for IHS. Although these IMMs do not
describe real particles since they do not interact according to a given
potential law, it must be stressed that several results derived from IMM \cite{S03,G03,GA05,SG07,MGV14} agree well with the predictions made from IHS. Moreover, in the framework of the
Boltzmann equation, Maxwell models can be introduced at the level of the cross section without any reference to a specific interaction potential \cite{E81,ETB06}. It is here noteworthy that some experiments \cite{KSSAOB05} for magnetic grains with dipolar interactions are well described by IMM.

One of the most widely studied states in granular gases is the so-called simple or uniform shear flow (USF) state. In the case of a binary granular mixture, it is characterized
by constant partial densities $n_r$ ($r=1,2$), a uniform granular temperature $T$, and a linear velocity profile $u_x=u_{1x}=u_{2x}=ay$ where $a$ is the constant shear rate and $\mathbf{u}_r$
denotes the mean velocity of species $r$. In this problem, the mass and heat fluxes vanish by symmetry and hence, the pressure tensor $\mathsf{P}$ is the relevant flux of the problem.
In the case of IMM, the elements of the pressure tensor were exactly determined \cite{GT10} in terms of the shear rate and the parameters of the mixture. Subsequently, the dynamics of
an impurity immersed in an inelastic Maxwell gas under USF was studied \cite{GT11,GT12a} by starting from the above exact solution \cite{GT10} [which holds for arbitrary concentration
$x_r=n_r/(n_1+n_2)$] and taking carefully the tracer limit (i.e, when the concentration of one of the species becomes negligible). Surprisingly, a non-equilibrium phase transition
was identified with a region (``ordered'' phase) where the contribution of impurities to the total kinetic energy is \emph{finite}. This unexpected behaviour was present
when the gas is sheared or when it evolves freely (namely, in the so-called homogeneous cooling state) \cite{GT12b}. In this latter case, we have recently analyzed \cite{GKT15}
the impact of this transition on the Navier-Stokes transport coefficients, showing that those coefficients exhibit a different dependence on the mass ratios and the coefficients
of restitution in the ``ordered'' and ``disordered'' phases.

The aim of this paper is to gauge the effect of the above non-equilibrium transition on the transport properties associated with impurities when the granular gas is shear flow driven.
As in the previous study \cite{GKT15} for the Navier-Stokes coefficients, in order to determine in a clean way the behaviour of the tracer transport coefficients in both non-equilibrium phases,
one has first to evaluate transport around USF for a general binary mixture (i.e., with $x_1\neq 0$) and then take the corresponding tracer limit ($x_1\to 0$). This first requires
the computation of the complete set of generalized transport coefficients of a granular binary mixture (with $x_1\neq 0$) in a state that deviates
from the USF by small spatial gradients. To get those coefficients, one should solve the set of coupled Boltzmann equations by means of a Chapman-Enskog-like method around the (local) shear
flow distributions for each species that retain all the hydrodynamic orders in the shear rate. This is the essential difference with respect to the conventional Chapman-Enskog method \cite{CC70}.
Since the base state (zeroth-order approximation) is anisotropic, tensorial quantities are required to describe the irreversible fluxes instead of scalar coefficients. The evaluation of these
generalized tensors has been recently carried out by the authors of the present paper \cite{GT15}. On the other hand, due to the technical difficulties involved in explicitly computing the
shear-rate dependence of the transport coefficients, only the mass transport of impurities was evaluated. This flux is characterized by the second-rank tensors $D_{ij}$ (diffusion tensor),
$D_{p,ij}$ (pressure diffusion tensor) and $D_{T,ij}$ (thermal diffusion tensor). Here, we will explicitly compute the above shear-rate dependent diffusion coefficients in the tracer limit.
The results show that the dependence of those coefficients on the shear rate and the parameters of the mixture is clearly different in both ordered and disordered phases.

The plan of the paper is as follows. In section \ref{sec2} we introduce the Boltzmann equation for IMM and present the USF problem. In addition, the tracer limit is also considered and the
ordered phases where impurities bear a finite contribution to the properties of the mixture are identified. Section \ref{sec3} deals with the description of the Chapman-Enskog-like method
to get the diffusion coefficients $D_{ij}$, $D_{p,ij}$ and $D_{T,ij}$. The algebraic equations defining those coefficients are explicitly written in section \ref{sec4} for arbitrary concentration.
Then, starting from the above general expressions, we derive their forms in the ordered and disordered phases when the tracer limit is considered. The dependence of the above coefficients on the
parameter space of the problem is illustrated in section \ref{sec5} for systems where impurities are lighter or heavier than the particles of the granular gas. Finally, we conclude in section \ref{sec6} with a brief discussion of the main findings of the paper.

\section{Inelastic Maxwell mixtures under shear flow}
\label{sec2}

\subsection{Boltzmann kinetic equation}

Let us consider a granular binary mixture modeled as an inelastic Maxwell model.
In the simplest version, the Boltzmann equation for IMM \cite{Maxwell}
can be obtained from the Boltzmann equation for IHS by replacing the
rate for collisions between particles of species $r$ and $s$ by an average velocity-independent collision rate. With this simplification and in the absence of external forces,
the set of nonlinear Boltzmann kinetic equations becomes
\begin{equation}
\label{2.1}
\left(\frac{\partial}{\partial t}+{\bf v}\cdot \nabla \right)f_{r}({\bf r},{\bf v};t)
=\sum_{s=1}^2\;J_{rs}\left[{\bf v}|f_{r}(t),f_{s}(t)\right] \;,
\end{equation}
where $f_{r}({\bf r},{\bf v},t)$ is the one-particle distribution function  of species $r$ ($r=1,2$) and the Boltzmann collision operator
$J_{rs}\left[{\bf v}_{1}|f_{r},f_{s}\right]$ for IMM describing the scattering of pairs of particles is
%\begin{eqnarray}
\beq
J_{rs}\left[{\bf v}_{1}|f_{r},f_{s}\right] =\frac{\omega_{rs}}{n_s\Omega_d}
\int \dd{\bf v}_{2}\int \dd\widehat{\boldsymbol {\sigma }}\left[ \alpha_{rs}^{-1}f_{r}({\bf r},{\bf v}_{1}',t)
f_{s}({\bf r},{\bf v}_{2}',t)
-f_{r}({\bf r},{\bf v}_{1},t)f_{s}({\bf r},{\bf v}_{2},t)\right]
\;.
\label{2.2}
\eeq
%\end{eqnarray}
In equation \eqref{2.2},
\begin{equation}
\label{2.3}
n_r=\int \dd {\bf v} f_r({\bf v})
\end{equation}
is the number density of species $r$, $\omega_{rs}$ is an effective collision frequency for collisions  of type $r$-$s$,  $\Omega_d=2\pi^{d/2}/\Gamma(d/2)$
is the total solid angle in $d$ dimensions, and $\alpha_{rs}\leq 1$ refers to the
constant coefficient of restitution  for collisions between particles of species $r$
with $s$.  In addition, the primes on the velocities denote the initial values $\{{\bf
v}_{1}^{\prime}, {\bf v}_{2}^{\prime}\}$ that lead to $\{{\bf v}_{1},{\bf v}_{2}\}$
following a binary collision:
\begin{equation}
\label{2.4}
{\bf v}_{1}^{\prime }={\bf v}_{1}-\mu_{sr}\left( 1+\alpha_{rs}
^{-1}\right)(\widehat{\boldsymbol {\sigma}}\cdot {\bf g}_{12})\widehat{\boldsymbol{\sigma}},
\eeq
\beq
{\bf v}_{2}^{\prime}={\bf v}_{2}+\mu_{rs}\left(
1+\alpha_{rs}^{-1}\right) (\widehat{\boldsymbol {\sigma}}\cdot {\bf
g}_{12})\widehat{\boldsymbol{\sigma}}\;,
\end{equation}
where ${\bf g}_{12}={\bf v}_1-{\bf v}_2$ is the relative velocity of the colliding pair,
$\widehat{\boldsymbol {\sigma}}$ is a unit vector directed along the centers of the two colliding spheres, and $\mu_{rs}=m_r/(m_r+m_s)$.

The effective collision frequencies $\omega_{rs}$ are independent of velocity but
depend in general on space an time through their dependence on density and temperature. As in previous works \cite{GT11,GT12a,GT12b}, we will consider a simple version of
IMM (``plain vanilla Maxwell model'') where one defines $\omega_{rs}$ as
\begin{equation}
\label{2.4.1}
\omega_{rs}=x_s\nu_0, \quad \nu_0=A n,
\end{equation}
where $x_s=n_s/n$ is the concentration or mole fraction of species $s$ and the value of the constant $A$ is irrelevant for our purposes. Here, $n=n_1+n_2$ is the total number
density of the mixture. The form of $\omega_{rs}$ is closer to the original model of Maxwell molecules for elastic mixtures \cite{GS03}. This plain vanilla model has been
previously employed by several authors \cite{MP02,BK02} and it is capable of capturing the essential physical effects in shearing problems \cite{G03,GT12a}.

At a hydrodynamic level, apart from the partial densities $n_r$, the relevant quantities in a binary mixture are the flow velocity  ${\bf u}$, and the ``granular'' temperature $T$. They are defined as 
\begin{equation}
\label{2.4.2}
\rho{\bf u}=\sum_r\rho_r{\bf u}_r=\sum_r\int \dd {\bf v}m_r{\bf v}f_r({\bf v}),
\end{equation}
\begin{equation}
\label{2.5}
nT=\sum_rn_rT_r=\sum_r\int \dd {\bf v}\frac{m_r}{d}V^2f_r({\bf v}),
\end{equation}
where $\rho_r=m_rn_r$, $\rho=\rho_1+\rho_2$ is
the total mass density, and ${\bf V}={\bf v}-{\bf u}$ is the peculiar velocity.
Apart from the hydrodynamic fields, an interesting quantity is the partial temperature $T_r$ of species $r$ defined as
\beq
\label{2.5.1}
n_r T_r=\int \dd {\bf v}\frac{m_r}{d}V^2f_r({\bf v}).
\eeq
The partial temperature $T_r$ measures the mean kinetic energy of species $r$. As confirmed by computer simulations \cite{computer}, experiments \cite{exp1,exp2} and kinetic theory calculations \cite{JM87,GD99}, the global granular temperature $T$ is in general different from the partial temperatures $T_r$. In addition,
the mass flux ${\bf j}_{r}$ of species $r$, the pressure tensor $\mathsf{P}$ and the heat flux $\mathbf{q}$ are given, respectively, by
\begin{equation}
{\bf j}_{r}=m_{r}\int \dd{\bf v}\,{\bf V}\,f_{r}({\bf v}),
\label{2.6}
\end{equation}
\begin{equation}
{\sf P}=\sum_{r}\,\int \dd{\bf v}\,m_{r}{\bf V}{\bf V}\,f_{r}({\bf  v}),
\label{2.7}
\end{equation}
\begin{equation}
{\bf q}=\sum_{r}\,\int \dd{\bf v}\,\frac{1}{2}m_{r}V^{2}{\bf V}
\,f_{r}({\bf v}).
\label{2.8}
\end{equation}
Finally, the rate of energy dissipated due to collisions among all species defines the (total) cooling rate $\zeta$ as
\begin{equation}
\sum_{r}\sum_{s}\; m_r\int \dd{\bf v}V^{2}J_{rs}[{\bf v}
|f_{r},f_{s}]=-d nT\zeta \;.
\label{2.9}
\end{equation}
Equation \eqref{2.7} also defines the partial contribution $\mathsf{P}_r$ of species $r$ to the total pressure tensor $\mathsf{P}$ as
\beq
\label{2.9.1}
\mathsf{P}_r=\int \dd{\bf v}\,m_{r}{\bf V}{\bf V}\,f_{r}({\bf v}).
\eeq
Note that $\mathbf{j}_1=-\mathbf{j}_2$ due to the definition \eqref{2.6}.

\subsection{Uniform shear flow}

Let us assume that the mixture is under USF. This state is
macroscopically characterized by constant densities, a uniform temperature, and a
linear velocity profile
\begin{equation}
\label{2.11}
{\bf u}(y)={\bf u}_1(y)={\bf u}_2(y)=ay \widehat{{\bold x}},
\end{equation}
where $a$ is the constant shear rate. This profile assumes no boundary
layer near the walls and is generated by the Lees-Edwards boundary conditions
\cite{LE72}, which are simply periodic boundary conditions in the local Lagrange frame
moving with the flow velocity \cite{DSBR86}. Thus, at a microscopic level, the velocity distribution functions $f_s$ of the USF state become \emph{uniform} when one refers the
velocity of the particles to the local Lagrangian frame moving at the flow velocity defined by equation \eqref{2.11}, i.e., $f_{s}({\bf r},{\bf v},t)=f_{s}({\bf V},t)$. In that case,
equation (\ref{2.1}) can be written as \cite{GS03}
\begin{equation}
\label{2.12}
\frac{\partial}{\partial t}f_1-aV_y\frac{\partial}{\partial V_x}f_1=J_{11}[f_1,f_1]+J_{12}[f_1,f_2].
\end{equation}
A similar equation holds for $f_2$.

Since $n_s$ and $T$ are uniform in the USF state, then the
mass and heat fluxes vanish and the pressure tensor is the only non-vanishing flux of the problem. Moreover, the only relevant balance equation is that
for the temperature. It can be obtained from equation (\ref{2.12}) and its
counterpart for species $2$; it is given by
\begin{equation}
\label{2.13}
\nu_0^{-1}\frac{\partial}{\partial t}\ln T=-\zeta^*-\frac{2a^*}{d} P_{xy}^*,
\end{equation}
where $\zeta^*\equiv \zeta/\nu_0$, $a^* \equiv a/\nu_0$, and $P_{xy}^* \equiv P_{xy}/p$. Here, $p=nT$ is the hydrostatic pressure. In the USF problem, the expression for $\zeta^*$ is \cite{G03}
\beq
\label{2.13.1}
\zeta^*=\frac{2}{d}\sum_r\sum_s\; x_rx_s
\mu_{sr}(1+\alpha_{rs})\left[\gamma_r-\frac{1+\alpha_{rs}}{2}(\gamma_r\mu_{sr}+\gamma_s\mu_{rs})
\right],
\eeq
where $\gamma_r\equiv T_r/T$ and use has been made of the property ${\bf j}_r={\bf 0}$. The {\em reduced} shear rate $a^*$ is the nonequilibrium relevant parameter of the USF
problem since it measures the distance of the system from the homogeneous cooling
state ($a^*=0$). According to equation (\ref{2.13}), the temperature changes in time
due to the competition of two opposite mechanisms: on the one hand, viscous heating
($-a^* P_{xy}^*>0$) and, on the other hand, energy dissipation in collisions ($-\zeta^*<0$). In general, since $a^*$ does not depend on time, there is no steady state unless $a^*$
takes the specific value given by the steady-state condition
\begin{equation}
\label{2.14}
a_s^*P_{s,xy}^*=-\frac{d}{2}\zeta_s^*,
\end{equation}
where $a_s^*$, $P_{s,xy}^*$ and $\zeta_s^*$ denotes the steady-state values of the (reduced) shear rate, the pressure tensor and the cooling rate, respectively. Beyond this particular
case, the (reduced) shear rate and the coefficients of restitution are not coupled and hence, one can study the combined effect on both quantities on the elements of the pressure tensor of the mixture.

The explicit forms of the (scaled) pressure tensors $P_{r,ij}^*=P_{r,ij}/p$ have been obtained in Ref.\ \cite{GT10} as nonlinear functions of the (reduced) shear rate, the coefficients of
restitution and the parameters of the mixture (masses and concentration). Their expressions are displayed in Appendix \ref{appA}. In particular, for long times, the temperature behaves as
\begin{equation}
\label{2.15}
T(t)=T(0)e^{\lambda \nu_0 t},
\end{equation}
where $\lambda$ is the largest root of a sixth-degree polynomial equation with coefficients depending on $a^*$, $\alpha_{rs}$, $x_1$ and the mass ratio $\mu\equiv m_1/m_2$. The results obtained
in Ref.\ \cite{GT12a} for $x_1 \neq 0$ show that, at a given value of $a^*$, the difference between the two largest roots of the above sixth-degree equation does not vanish. This means that the
asymptotic time dependence of the partial pressure tensors $P_{r,ij}^*$ is always ruled by one of the roots.

\subsection{Tracer limit ($x_1 \to 0$)}

We assume now that the concentration of one of the species (say for instance, species 1) becomes negligible. In the tracer limit ($x_1\to 0$), the sixth-degree equation for $\lambda$ factorizes
into two cubic equations with the following largest roots:
\beq
\label{2.16}
\lambda_2^{(0)}=
\frac{(1+\alpha_{22})^2}{d+2}F(\widetilde{a})-\frac{1-\alpha_{22}^2}{2d},
\eeq
\beq
\label{2.19}
\lambda_1^{(0)}=
\frac{2\mu_{21}^2}{d+2}(1+\alpha_{12})^2F\left(\frac{\widetilde{a}}{2\mu_{21}^2}
\frac{(1+\alpha_{22})^2}{(1+\alpha_{12})^2}\right)
-\frac{2}{d}\mu_{21}(1+\alpha_{12})\left[1-\frac{\mu_{21}}{2}(1+\alpha_{12})\right],
\eeq
where
\begin{equation}
\label{2.17}
F(x)\equiv \frac{2}{3}\sinh^2\left[\frac{1}{6}\cosh^{-1}\left(1+\frac{27}{d}x^2\right)\right]
\end{equation}
and
\begin{equation}
\label{2.18}
\widetilde{a}=\frac{2(d+2)}{(1+\alpha_{22})^2}
a^*.
\end{equation}
The root $\lambda_2^{(0)}$ rules the dynamics of the host fluid (excess component)
while the evolution of the tracer species is governed by $\lambda_1^{(0)}$.

As said before, the largest of all roots, $\lambda_\text{max}$, is the relevant
one to obtain the asymptotic values of the (scaled) pressure tensors $\mathsf{P}_r^*$. In particular, the energy ratio $E_1/E=x_1\gamma_1$ (or equivalently, the reduced partial pressure
$p_1^*=n_1T_1/p$) can be easily obtained from the pressure tensor $\mathsf{P}_1^*$ associated with the tracer particles. It was shown in Refs.\ \cite{GT11,GT12a} that the behaviour of
the system is qualitatively very different depending on $\lambda_\text{max}=\lambda_1^{(0)}$ or $\lambda_\text{max}= \lambda_2^{(0)}$. Thus, when $\lambda_2^{(0)}>\lambda_1^{(0)}$,
$E_1/E=0$ when $x_1\to 0$ as expected and $T_1/T_2\equiv \text{finite}$. This region of the parameter space is coined as the ``disordered'' phase.

On the other hand, if $\lambda_1^{(0)}>\lambda_2^{(0)}$, then $T_1/T_2\to \infty$ but surprisingly $E_1/E \neq 0$. We found two different families of ``ordered'' phase:
\begin{itemize}
\item A light impurity phase which is present when $a^*>a_c^*(\mu,\alpha_{rs})$ and $\mu<\mu_{\text{th}}^{(-)}$ where
\begin{equation}
\label{2.18.3}
\mu_{\text{th}}^{(-)} \,=\, \sqrt{2}
\frac{1+\alpha_{12}}{1+\alpha_{22}}-1.
\end{equation}
This phase can also be observed at vanishing shear rate ($a^*=0$) when the mass ratio $\mu>\mu_{\text{HCS}}^{(+)}$ or $\mu<\mu_{\text{HCS}}^{(-)}$ where \cite{GT12b}
\begin{equation}
\label{eq:muhcs}
\mu_{\text{HCS}}^{(-)}=\frac{\alpha_{12}-\sqrt{\frac{1+\alpha_{22}^2}{2}}}
{1+\sqrt{\frac{1+\alpha_{22}^2}{2}}}, \quad
\mu_{\text{HCS}}^{(+)}=\frac{\alpha_{12}+\sqrt{\frac{1+\alpha_{22}^2}{2}}}
{1-\sqrt{\frac{1+\alpha_{22}^2}{2}}}.
\end{equation}
Note that while the upper bound $\mu_{\text{HCS}}^{(+)}$ is well defined for all values of $\al_{12}$ and $\al_{22}$, the lower one is only positive when $\al_{12}>\sqrt{(1+\al_{22})^2/2}$
(asymmetric dissipation).

\item A heavy impurity phase ($\mu>\mu_{\text{HCS}}^{(+)}$), which cannot accommodate large shear rates and requires $a^*<a^{*(+)}$, where
\begin{equation}
a^{*(+)} \,=\, \frac{1+d-\alpha_{22}}{d} \,
\sqrt{\frac{1-\alpha_{22}^2}{2(d+2)}}.
\label{2.18.2}
\end{equation}
\end{itemize}
The existence of the light impurity ordered phase was already found years ago for elastic collisions \cite{MSG96}. The explicit form of $E_1/E$ is provided \cite{note} in the
Appendix C of Ref.\ \cite{GT12a}. Note that the above results do not depend of the impurity-impurity coefficient of restitution $\al_{11}$, which is intuitively expected.

As alluded to in the Introduction, the goal here is to analyze the fingerprint of this nonequilibrium transition on the diffusion coefficients associated to the tracer species. In order to do it,
we have to determine them first for \emph{arbitrary} $x_1$. This will be carried out in the next Section by solving the Boltzmann equation of the mixture by means of a Chapman-Enskog-like expansion.

\section{Chapman-Enskog-like expansion around USF}
\label{sec3}

We assume now that we excite the USF by small spatial perturbations, in order to get the diffusion transport coefficients associated with the mass flux.
We start from the set of Boltzmann equations
(\ref{2.1}) with a general time and space dependence. Let $u_{0,i}=a_{ij} r_j$ be the flow velocity of the {\em undisturbed} USF state, where $a_{ij}=a\delta_{ix}\delta_{jy}$.
In the {\em disturbed} state however, the true velocity ${\bf u}$ is in general
different from $\mathbf{u}_0$ \cite{L06,G06,G07}, i.e., $u_i=u_{0,i}+\delta u_i$, $\delta u_i$
being a small perturbation to $u_{0,i}$. Thus, in the perturbed USF state, the peculiar velocity 
is $\mathbf{c}=\mathbf{V}-\delta \mathbf{u}$, where
$\mathbf{V}=\mathbf{v}-\mathbf{u}_0$. In the Lagrangian frame moving with ${\bf u}_0$, the Boltzmann equations (\ref{2.1}) reads
\begin{subequations}
\begin{equation}
\label{3.1} \frac{\partial}{\partial t}f_1-aV_y\frac{\partial}{\partial
V_x}f_1+\left({\bf V}+{\bf u}_0\right)\cdot \nabla f_1=J_{11}[f_1,f_1]+J_{12}[f_1,f_2],
\end{equation}
\begin{equation}
\label{3.2} \frac{\partial}{\partial t}f_2-aV_y\frac{\partial}{\partial
V_x}f_2+\left({\bf V}+{\bf u}_0\right)\cdot \nabla f_2
=J_{22}[f_2,f_2]+J_{21}[f_2,f_1],
\end{equation}
\end{subequations}
where the derivative $\nabla f_r$ is taken at constant ${\bf V}$. The macroscopic
balance equations follow as
\begin{equation}
\label{3.3}
\partial_t n_r+\mathbf{u}_0\cdot \nabla n_r+\nabla \cdot (n_r\delta \mathbf{u})=-
\frac{\nabla \cdot \mathbf{j}_r}{m_r}, \quad (r=1,2),
\end{equation}
\begin{equation}
\label{3.4}
\partial_t\delta u_i+a_{ij}\delta u_j+({\bf u}_0+\delta {\bf u})\cdot \nabla \delta u_i=-\rho^{-1}\nabla_j P_{ij},
\end{equation}
\beq
\label{3.5}
\frac{d}{2}n\partial_tT+\frac{d}{2}n({\bf u}_0+\delta {\bf u})\cdot \nabla
T=-aP_{xy}-\frac{d}{2}T\sum_{r}\frac{\nabla \cdot {\bf j}_r}{m_r}-
\left(\nabla \cdot {\bf q}+{\sf P}:\nabla
\delta {\bf u}+\frac{d}{2}p\zeta\right),
\eeq
where the mass flux $\mathbf{j}_r$, the pressure tensor $\mathsf{P}$, the heat flux 
$\mathbf{q}$, and the cooling rate $\zeta$ are defined by equations (\ref{2.6}), (\ref{2.7}), (\ref{2.8}),
and (\ref{2.9}), respectively, with the replacement $\mathbf{V}\rightarrow \mathbf{c}$.

The deviations from the USF state are assumed small; the spatial gradients of the hydrodynamic fields are thus small as well. Here, as in previous works on granular mixtures \cite{GD02},
we chose the mole fraction $x_1$, the pressure $p$, the temperature $T$, and the local flow velocity $\delta {\bf u}$ as the relevant hydrodynamic fields. Since the system is strongly sheared,
a solution to the set of Boltzmann equations (\ref{3.1}) and (\ref{3.2}) can be obtained by means of a generalization of the conventional Chapman-Enskog method \cite{CC70} in which the
velocity distribution function is expanded around a {\em local} shear flow reference state in terms of the small spatial gradients of the hydrodynamic fields relative to those of USF.
This is the main new ingredient of the expansion.

This type of Chapman-Enskog-like expansion has been already considered to get the set of shear-rate dependent transport coefficients for monodisperse systems in the case of inelastic hard spheres \cite{L06,G06} and inelastic
Maxwell models \cite{G07}. More recently, the method has been extended to the case of granular mixtures \cite{GT15}. Since the procedure involved in the evaluation of the first-order approximation to the mass flux (which is the quantity needed to analyze the diffusion coefficients) has been widely exposed in Ref.\ \cite{GT15}, we will start here our study on tracer diffusion coefficients by adapting the results derived in this paper \cite{GT15} to the special vanilla Maxwell model [see equation \eqref{2.4.1}]. More technical details on the application of the Chapman-Enskog-like method can be found in the latter reference.

\subsection{First-order approximation to the mass flux}

To first order in the gradients, the mass flux ${\bf j}_1^{(1)}$ of species $1$ is given by
\begin{equation}
\label{3.24}
j_{1,i}^{(1)}=-\frac{m_1m_2n}{\rho}D_{ij}\frac{\partial x_1}{\partial r_j}-
\frac{\rho}{p}D_{p,ij}\frac{\partial p}{\partial r_j}-\frac{\rho}{T}D_{T,ij}\frac{\partial T}{\partial r_j},
\quad j_{2,i}^{(1)}=-j_{1,i}^{(1)}.
\end{equation}
The diffusion tensors $D_{ij}$, $D_{p,ij}$, and $D_{T,ij}$ are defined as
\begin{equation}
\label{3.25} 
D_{ij}=-\frac{\rho}{n m_2}\int \dd{\bf c}\,c_i\;{\cal A}_{1,j}({\bf c}),
\end{equation}
\begin{equation}
\label{3.26} 
D_{p,ij}=-\frac{pm_1}{\rho}\int \dd{\bf c}\,c_i\;{\cal B}_{1,j}({\bf c}),
\end{equation}
\begin{equation}
\label{3.27} 
D_{T,ij}=-\frac{Tm_1}{\rho}\int \dd{\bf c}\,c_i\;{\cal C}_{1,j}({\bf c}),
\end{equation}
where ${\boldsymbol {\cal A}}_{1}(\mathbf{c})$, ${\boldsymbol {\cal B}}_{1}(\mathbf{c})$ and ${\boldsymbol {\cal C}}_{1}(\mathbf{c})$ are the solutions of the following set of linear integral equations:
\begin{widetext}
\begin{eqnarray}
\label{3.18} \lambda \nu_0\left(p\partial_p+T\partial_T\right){\boldsymbol {\cal A}}_{1}-& & a
c_y\frac{\partial}{\partial c_x}{\boldsymbol {\cal A}}_{1}+{\cal L}_1 {\boldsymbol {\cal A}}_{1}+
{\cal M}_1 {\boldsymbol {\cal A}}_{2}={\bf A}_{1}\nonumber\\
& & +\left(\frac{2a}{d}\frac{\partial P_{xy}^{(0)}}{\partial x_1}+p \frac{\partial
\zeta^{(0)}}{\partial x_1}\right) {\boldsymbol {\cal B}}_{1} +\left(\frac{2aT}{d
p}\frac{\partial P_{xy}^{(0)}}{\partial x_1}+T \frac{\partial \zeta^{(0)}}{\partial
x_1}\right){\boldsymbol {\cal C}}_{1},
\end{eqnarray}
\begin{eqnarray}
\label{3.19} \lambda \nu_0\left(p\partial_p+T\partial_T\right){\boldsymbol {\cal
B}}_{1}-& &\left[ \frac{2a}{d}\partial_p P_{xy}^{(0)}+(1+p\partial_p)\zeta^{(0)}+ a
c_y\frac{\partial}{\partial c_x}\right]{\boldsymbol {\cal B}}_{1}+{\cal L}_1
{\boldsymbol {\cal B}}_{1}+
{\cal M}_1 {\boldsymbol {\cal B}}_{2}={\bf B}_{1}\nonumber\\
& & -\left[\frac{2aT}{d p^2}\left(1-p\partial_p \right)P_{xy}^{(0)}-
\frac{T}{p}\partial_p\zeta^{(0)}\right] {\boldsymbol {\cal C}}_{1},
\end{eqnarray}
\begin{eqnarray}
\label{3.20} \lambda \nu_0\left(p\partial_p+T\partial_T\right){\boldsymbol {\cal
C}}_{1}-& &\left[ \left(1+T\partial_T \right)\left(\frac{2a}{d
p}P_{xy}^{(0)}+\zeta^{(0)}\right)+ a c_y\frac{\partial}{\partial
c_x}\right]{\boldsymbol {\cal C}}_{1}+{\cal L}_1 {\boldsymbol {\cal C}}_{1}+
{\cal M}_1 {\boldsymbol {\cal C}}_{2}={\bf C}_{1}\nonumber\\
& & +\left[\partial_T\left(\frac{2a}{d}P_{xy}^{(0)}+p\zeta^{(0)}\right)\right]
{\boldsymbol {\cal B}}_{1}.
\end{eqnarray}
\end{widetext}
Here, we have introduced the quantities 
\begin{equation}
\label{b3} A_{1,i}({\bf c})=-\frac{\partial f_1^{(0)}}{\partial x_1}c_i-\frac{1}{\rho}
\frac{\partial f_1^{(0)}}{\partial c_j}\frac{\partial P_{ij}^{(0)}}{\partial
x_1},
\end{equation}
\begin{equation}
\label{b4} B_{1,i}({\bf c})=-\frac{\partial f_1^{(0)}}{\partial p} c_i-\frac{1}{\rho}
\frac{\partial f_1^{(0)}}{\partial c_j}\frac{\partial P_{ij}^{(0)}}{\partial p},
\end{equation}
\begin{equation}
\label{b5} C_{1,i}({\bf c})=-\frac{\partial f_1^{(0)}}{\partial T} c_i-\frac{1}{\rho}
\frac{\partial f_1^{(0)}}{\partial c_j}\frac{\partial P_{ij}^{(0)}}{\partial T}.
\end{equation}
Moreover, ${\cal L}_1$ and ${\cal M}_1$ are the linearized Boltzmann collision operators around the reference USF state:
\begin{subequations}
\begin{equation}
\label{3.22} {\cal L}_1X=-\left(J_{11}[f_1^{(0)},X]+J_{11}[X,f_1^{(0)}]+J_{12}[X,f_2^{(0)}]\right),
\end{equation}
\begin{equation}
\label{3.23} {\cal M}_1X=-J_{12}[f_2^{(0)},X].
\end{equation}
\end{subequations}
In equations \eqref{3.18}--\eqref{3.20}, $\zeta^{(0)}$ and $P_{r,ij}^{(0)}$ are the zeroth-order approximations to the cooling rate and the partial pressure tensor, respectively,
and $f_1^{(0)}$ is the zeroth-order distribution function.

\section{Shear-rate dependent diffusion coefficients. Tracer limit}
\label{sec4}

The generalized diffusion coefficients $D_{ij}$, $D_{p,ij}$, and $D_{T,ij}$ are nonlinear functions of the shear rate and the parameters of the mixture (masses, concentration
and coefficients of restitution). In dimensionless form, the above coefficients can be written as
\beq
\label{n1}
D_{ij}=\frac{\rho T}{m_1m_2\nu_0}D_{ij}^*,
\eeq
\beq
\label{n2}
D_{p,ij}=\frac{p}{\rho \nu_0}D_{p,ij}^*, \quad
D_{T,ij}=\frac{p}{\rho \nu_0}D_{T,ij}^*.
\eeq
In order to determine them, one has to multiply equations \eqref{3.18}--\eqref{3.20} by $m_1 c_j$ and integrate over ${\bf c}$. After some algebra, the (scaled) diffusion coefficients
$D_{ij}^*$, $D_{p,ij}^*$ and $D_{T,ij}^*$ obey the following set of coupled algebraic equations:
\begin{widetext}
\begin{equation}
\label{4.11} -\left(\lambda+\nu_D^*\right)D_{ij}^*-a_{ik}^*D_{kj}^*=
\frac{\rho_1}{\rho}\frac{\partial P_{ij}^{*}}{\partial x_1}-\frac{\partial P_{1,ij}^{*}}{\partial x_1}+\frac{\partial \lambda}{\partial x_1}\left(D_{p,ij}^*+D_{T,ij}^*\right),
\end{equation}
\begin{equation}
\label{4.12}
\left(2\lambda-a^*\frac{\partial \lambda}{\partial a^*}+\nu_D^*\right)D_{p,ij}^*
+a_{ik}^*D_{p,kj}^*=-\left(\frac{\rho_1}{\rho}P_{ij}^*-P_{1,ij}^*\right)+\frac{\rho_1}{\rho}
a^*\frac{\partial P_{ij}^*}{\partial a^*}-a^*\frac{\partial P_{1,ij}^*}{\partial a^*}
-\left(\lambda-a^*\frac{\partial \lambda}{\partial a^*}\right)D_{T,ij}^*,
\end{equation}
\begin{equation}
\label{4.13}
\left(a^*\frac{\partial \lambda}{\partial a^*}+\nu_D^*\right)D_{T,ij}^*
+a_{ik}^*D_{T,kj}^*=-\frac{\rho_1}{\rho}a^*\frac{\partial P_{ij}^*}{\partial a^*}+a^*
\frac{\partial P_{1,ij}^*}{\partial a^*}+\left(\lambda-a^*\frac{\partial \lambda}{\partial a^*}\right)D_{p,ij}^*.
\end{equation}
\end{widetext}
In equations \eqref{4.11}--\eqref{4.13}, we have introduced the dimensionless quantities $\zeta^*\equiv \zeta^{(0)}/\nu_0$, $P_{r,ij}^*\equiv P_{r,ij}^{(0)}/p$,
$P_{ij}^*\equiv P_{ij}^{(0)}/p=P_{1,ij}^*+P_{2,ij}^*$ and
\begin{equation}
\label{4.4}
\nu_D^*=\frac{\rho \omega_{12}^*}{d\rho_2}\mu_{21}(1+\alpha_{12}),
\end{equation}
where $\omega_{12}^*\equiv \omega_{12}/\nu_0$. Upon deriving equations \eqref{4.11}--\eqref{4.13} use has been made of the result
\beq
\label{derlamba}
a^*\frac{\partial \lambda}{\partial a^*}=-\frac{2a^*}{d}
\left(P_{xy}^*+a^*\frac{\partial P_{xy}^*}{\partial a^*}\right)-a^*\frac{\partial \zeta^*}{\partial a^*},
\eeq
that comes from the identity
\beq
\label{2.15.1}
\lambda=-\left(\zeta^*+\frac{2a^*}{d} P_{xy}^*\right).
\eeq

The solution to equations \eqref{4.11}--\eqref{4.13} provides the explicit forms of the set of diffusion coefficients for arbitrary concentration. In particular, in the absence of
shear field ($a^*=0$), $P_{ij}^*=\delta_{ij}$, $P_{r,ij}^*=x_r \gamma_r\delta_{ij}$, and so the tensorial quantities $D_{ij}^*$, $D_{p,ij}^*$ and $D_{T,ij}^*$ becomes scalar
coefficients, namely,  $D_{ij}^*=D^*\delta_{ij}$, $D_{p,ij}^*=D_p^*\delta_{ij}$ and $D_{T,ij}^*=D_T^*\delta_{ij}$ where
\begin{equation}
\label{4.14}
D^*=\left(\nu_D^*+\lambda\right)^{-1}\left[\frac{\partial}{\partial x_1}(x_1\gamma_1)-
\frac{\partial \lambda}{\partial x_1}\left(D_p^*+D_T^*\right)\right],
\end{equation}
\begin{equation}
\label{4.15}
D_p^*=x_1\gamma_1\left(1-\frac{pm_1}{\rho T_1}\right)\left(\nu_D^*+2\lambda+
\frac{\lambda^{2}}{\nu_D^*}\right)^{-1},
\end{equation}
\begin{equation}
\label{4.16}
D_T^*=\frac{\lambda}{\nu_D^*}D_p^*.
\end{equation}
The expressions \eqref{4.14}--\eqref{4.16} are consistent with those previously derived in the Navier-Stokes hydrodynamic order \cite{GA05}.

We now address the tracer limit ($x_1 \to 0$) for $D_{ij}^*$, $D_{p,ij}^*$ and $D_{T,ij}^*$. The analysis is quite delicate and shows
that the above coefficients turn out to be qualitatively different in the disordered and ordered phase, as may have been expected from the previous results obtained in the
Navier-Stokes order \cite{GKT15}. Let us consider each phase separately.

\subsection{Disordered phase}

In the disordered phase, $\lambda=\lambda_2^{(0)}$, the temperature ratio is finite and the energy ratio $p_1^*=0$. Moreover, the results displayed in Appendix \ref{appA} show that
in the disordered phase $P_{1xy,\text{dis}}^{*}$ and $P_{1yy,\text{dis}}^{*}$ are proportional to $x_1$ and hence, they vanish in the tracer limit. On the other hand,
\begin{equation}
\label{5.1}
\lim_{x_1\to 0}\left(\frac{\partial P_{1ij,\text{dis}}^{*}}{\partial x_1}\right)_{p,T}={\cal P}_{1ij,\text{dis}}^{(1)},
\end{equation}
where the explicit forms of the relevant elements ${\cal P}_{1xy,\text{dis}}^{(1)}$ and ${\cal P}_{1yy,\text{dis}}^{(1)}$ are defined by equations \eqref{a14.1} and \eqref{a14.2}, respectively.
In these conditions, the set of equations \eqref{4.12} and \eqref{4.13} obeyed by the tensors $D_{p,ij}^*$ and $D_{T,ij}^*$ become a set of \emph{homogeneous} equations whose solution yields
$D_{p,ij}^*=D_{T,ij}^*=0$. In the case of the diffusion tensor $D_{ij}^*$, equation \eqref{4.11} becomes
\begin{equation}
\label{5.2} \left(\lambda_2^{(0)}+\nu_D^*\right)D_{ij}^*+a_{ik}^*D_{kj}^*=
{\cal P}_{1ij,\text{dis}}^{(1)},
\end{equation}
whose solution is
\begin{equation}
\label{5.3}
D_{ij}^*=\frac{1}{\lambda_2^{(0)}+\nu_D^*}\left(\delta_{ik}-\frac{a_{ik}^*}
{\lambda_2^{(0)}+\nu_D^*}\right)
{\cal P}_{1kj,\text{dis}}^{(1)}.
\end{equation}
Here, we have introduced the tensor $a_{k\ell}^*=a^*\delta_{kx}\delta_{\ell y}$ and $\nu_D^*=\mu_{21}(1+\al_{12})/2$ in the tracer limit. Equation \eqref{5.3} was already obtained in
Ref.\ \cite{G03} in the study of diffusion of impurities in a sheared inelastic Maxwell gas. Moreover, when $a^*=0$,
${\cal P}_{1ij,\text{dis}}^{(1)}=\gamma_1 \delta_{ij}$ and one recovers the results derived in the Navier-Stokes approximation \cite{GA05}.

\subsection{Ordered phase}

The calculations in the ordered phase are, expectedly, more intricate. In this case, $\lambda=\lambda_1^{(0)}$, $\gamma_1 \to \infty$ but $p_1^{(0)}\equiv E_1/E\neq 0$. Here, $p_1^{(0)}$ is the
zeroth-order contribution to the expansion of $p_1^*$ in powers of the concentration $x_1$, i.e.,
\beq
\label{p1}
p_1^*=p_1^{(0)}+p_1^{(1)}x_1+\ldots.
\eeq
In addition, in order to obtain the diffusion tensors, we need also to evaluate the two first terms of the expansion of the tracer pressure tensor $P_{1ij,\text{ord}}^{*}$ in powers of $x_1$:
\beq
\label{P1}
P_{1ij,\text{ord}}^{*}={\cal P}_{1ij,\text{ord}}^{(0)}+
{\cal P}_{1ij,\text{ord}}^{(1)}x_1+\ldots.
\eeq
The explicit expressions of ${\cal P}_{1ij,\text{ord}}^{(0)}$ and ${\cal P}_{1ij,\text{ord}}^{(1)}$ are provided in
Appendix \ref{appA}. Once these quantities are known, the set of coupled equations verified by the diffusion tensors in the ordered phase can be obtained after taking the tracer limit in
equations \eqref{4.11}--\eqref{4.13}. The result is
\beq
\label{5.5}
-\left(\lambda_1^{(0)}+\nu_D^*\right)D_{ij}^*-a_{ik}^*D_{kj}^*=-
{\cal P}_{1ij,\text{ord}}^{(1)} +\lambda_1^{(1)}\left(D_{p,ij}^*+D_{T,ij}^*\right),
\eeq
\beq
\label{5.6}
\left(2\lambda_1^{(0)}-a^*\frac{\partial \lambda_1^{(0)}}{\partial a^*}+\nu_D^*\right)D_{p,ij}^*
+a_{ik}^*D_{p,kj}^*={\cal P}_{1ij,\text{ord}}^{(0)}
-a^*\frac{\partial
{\cal P}_{1ij,\text{ord}}^{(0)}}{\partial a^*}
-\left(\lambda_1^{(0)}-a^*\frac{\partial \lambda_1^{(0)}}{\partial a^*}\right)D_{T,ij}^*,
\eeq
\beq
\label{5.7}
\left(2\lambda_1^{(0)}-a^*\frac{\partial \lambda_1^{(0)}}{\partial a^*}+\nu_D^*\right)D_{T,ij}^*
+a_{ik}^*D_{T,kj}^*=a^*\frac{\partial
{\cal P}_{1ij,\text{ord}}^{(0)}}{\partial a^*}
+\left(\lambda_1^{(0)}-a^*\frac{\partial \lambda_1^{(0)}}{\partial a^*}\right)D_{p,ij}^*,
\eeq
where the explicit expression of $\lambda_1^{(1)}$ is provided in Appendix C of Ref.\ \cite{GT12a}. Equations \eqref{5.5}--\eqref{5.7} are the most relevant results of the present paper
since their solution provides the dependence of $D_{ij}^*$, $D_{p,ij}^*$ and $D_{T,ij}^*$ on the parameters of the system in the ordered phase. In particular, once the set of coupled
algebraic equations \eqref{5.6} and \eqref{5.7} for $D_{p,ij}^*$ and $D_{T,ij}^*$ are solved, the solution to equation \eqref{5.5} is simply
\beq
\label{5.8}
D_{ij}^*=\frac{1}{\lambda_1^{(0)}+\nu_D^*}\left(\delta_{ik}-\frac{a_{ik}^*}
{\lambda_1^{(0)}+\nu_D^*}\right)\left[{\cal P}_{1kj,\text{ord}}^{(1)}-
\lambda_1^{(1)}\left(D_{p,kj}^*+D_{T,kj}^*\right)\right].
\eeq

\begin{figure}
%[hbtp]
%\begin{center}
%\begin{tabular}{lr}
%\resizebox{3.7cm}{!}
{\includegraphics[width=0.35\columnwidth]{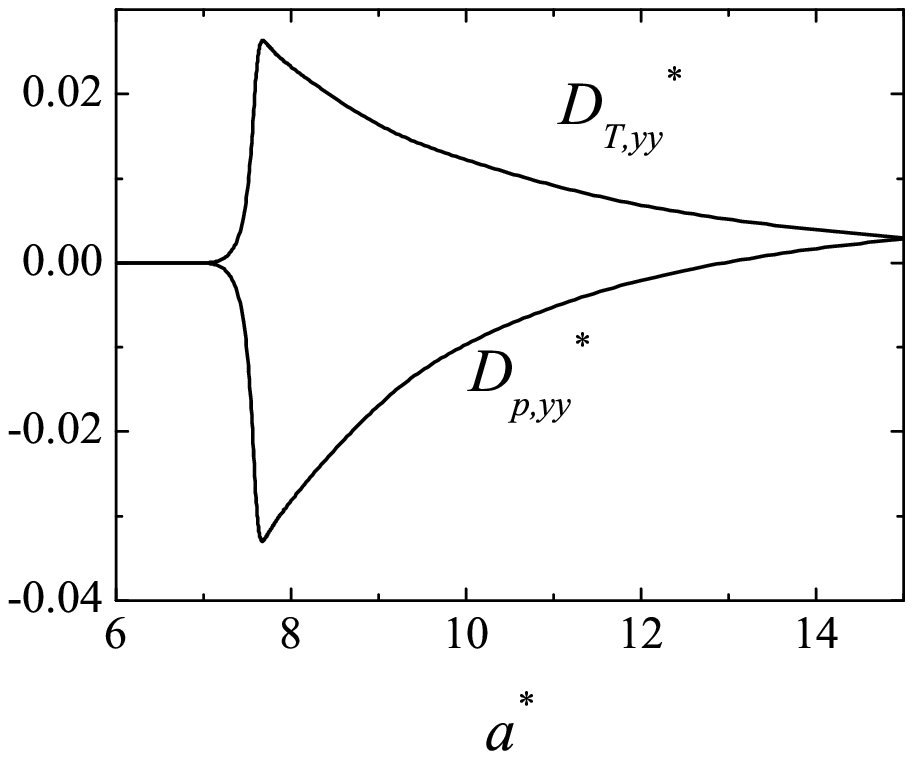}}
%&\resizebox{3.2cm}{!}
{\includegraphics[width=0.4\columnwidth]{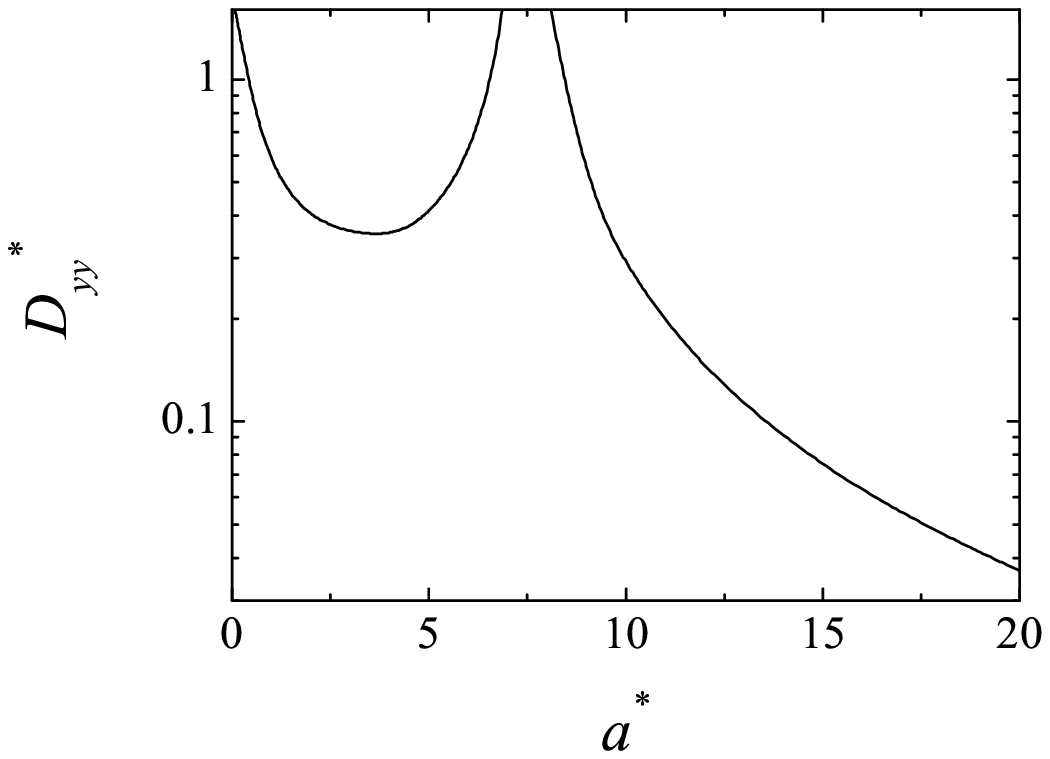}}
%\end{tabular}
%\resizebox{6.5cm}{!}{\includegraphics{fig4.eps}}
%\end{center}
\caption{Shear rate dependence of the diffusion coefficients $D_{p,yy}^*$, $D_{T,yy}^*$ and $D_{yy}^*$ for a three-dimensional system with a mass ratio $\mu=0.2$ and a (common) coefficient of
restitution $\al=\al_{22}=\al_{12}=0.9$. In this case, the value of the critical shear rate $a_c^*$ beyond which the ordered phase appears is $a_c^* \simeq 7.56$.
\label{fig1}}
\end{figure}
\begin{figure}
%[hbtp]
%\begin{center}
%\begin{tabular}{lr}
%\resizebox{3.7cm}{!}
{\includegraphics[width=0.35\columnwidth]{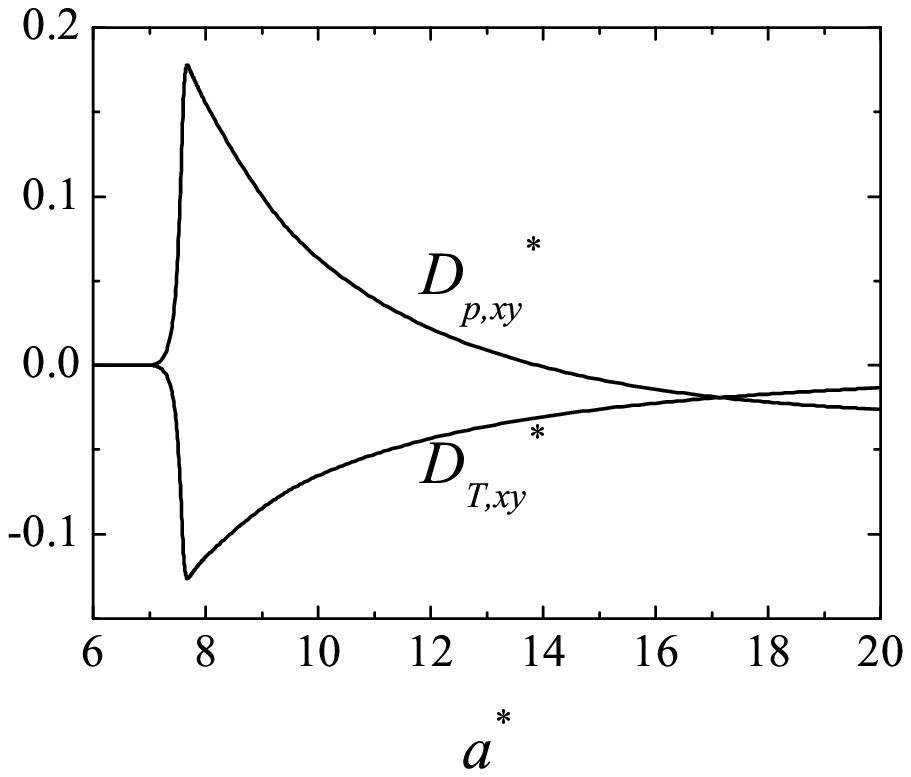}}
%&\resizebox{3.2cm}{!}
{\includegraphics[width=0.4\columnwidth]{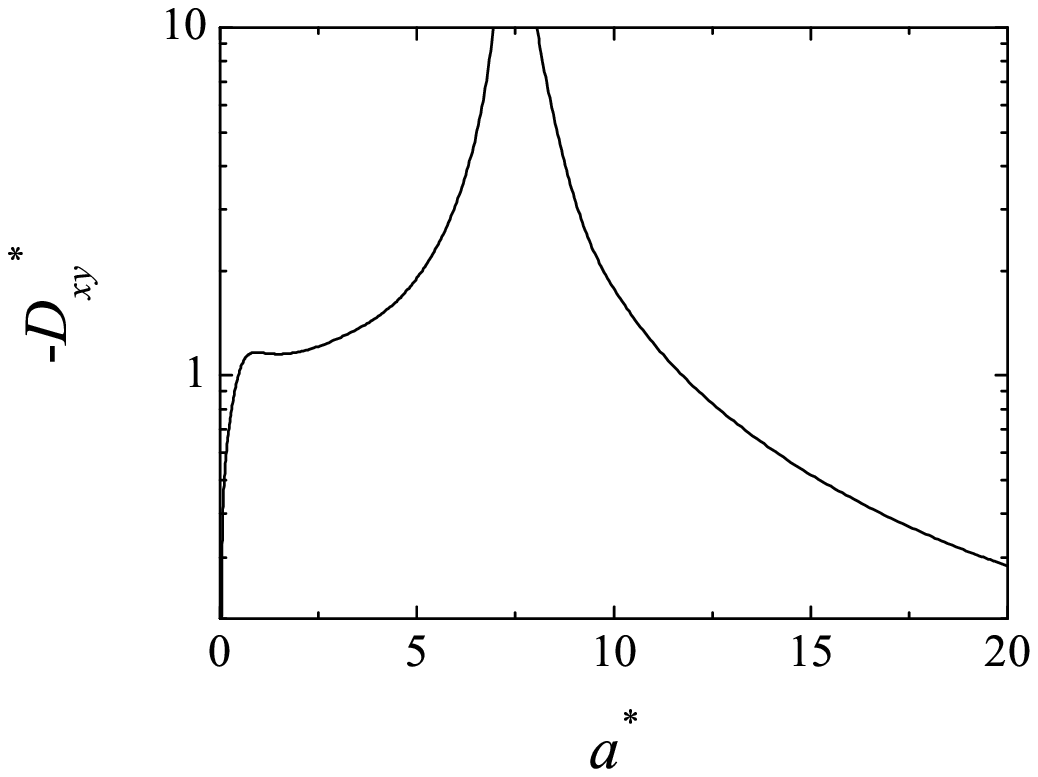}}
%\end{tabular}
%\resizebox{6.5cm}{!}{\includegraphics{fig4.eps}}
%\end{center}
\caption{Same as in Fig.\ \ref{fig1} for the coefficients $D_{p,xy}^*$, $D_{T,xy}^*$ and $-D_{xy}^*$.
\label{fig2}}
\end{figure}
\begin{figure}
%[hbtp]
%\begin{center}
%\begin{tabular}{lr}
%\resizebox{3.7cm}{!}
{\includegraphics[width=0.35\columnwidth]{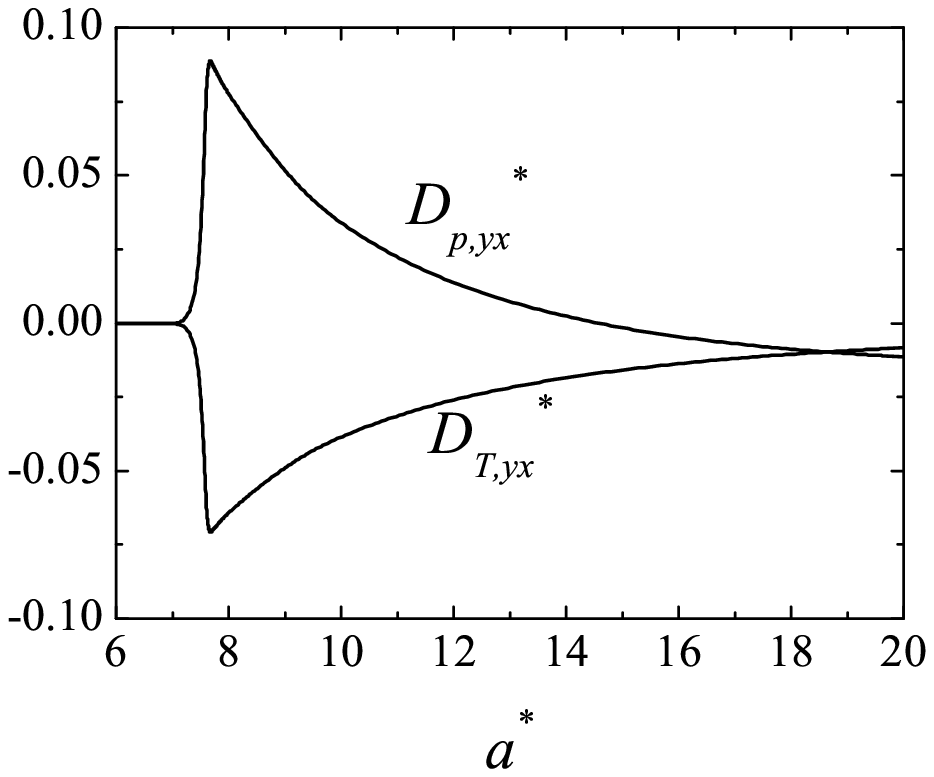}}
%&\resizebox{3.2cm}{!}
{\includegraphics[width=0.4\columnwidth]{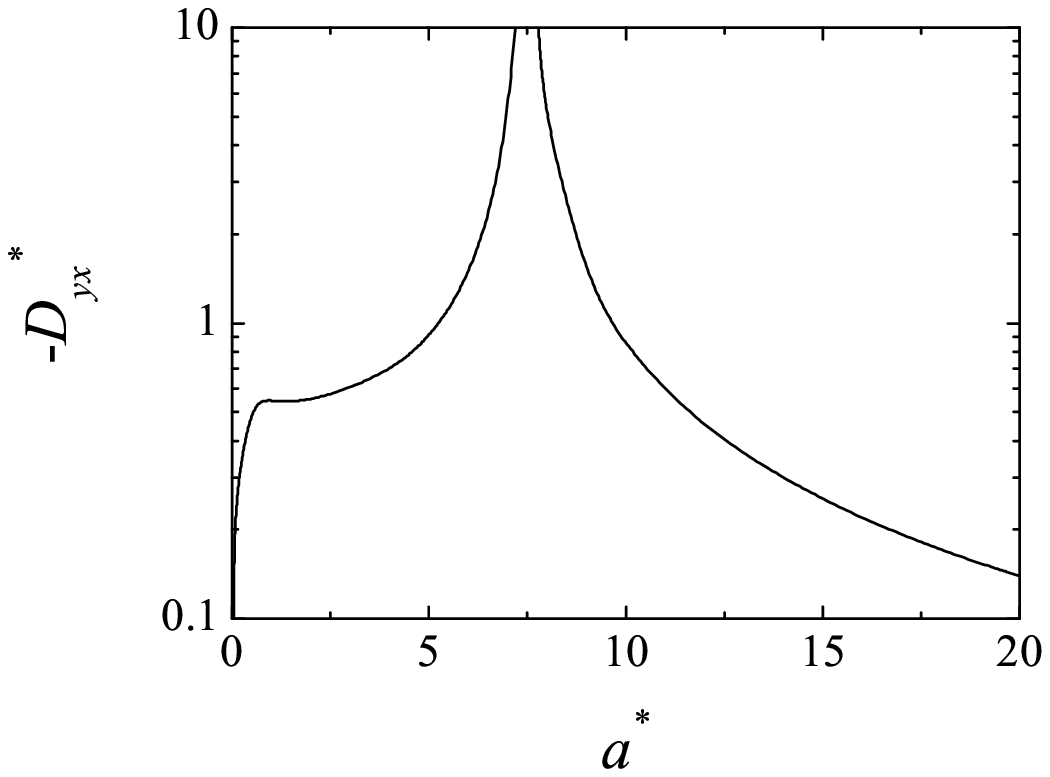}}
%\end{tabular}
%\resizebox{6.5cm}{!}{\includegraphics{fig4.eps}}
%\end{center}
\caption{Same as in Fig.\ \ref{fig1} but for the coefficients $D_{p,yx}^*$, $D_{T,yx}^*$ and $-D_{yx}^*$.
\label{fig3}}
\end{figure}
\begin{figure}
%[hbtp]
%\begin{center}
%\begin{tabular}{lr}
%\resizebox{3.7cm}{!}
{\includegraphics[width=0.35\columnwidth]{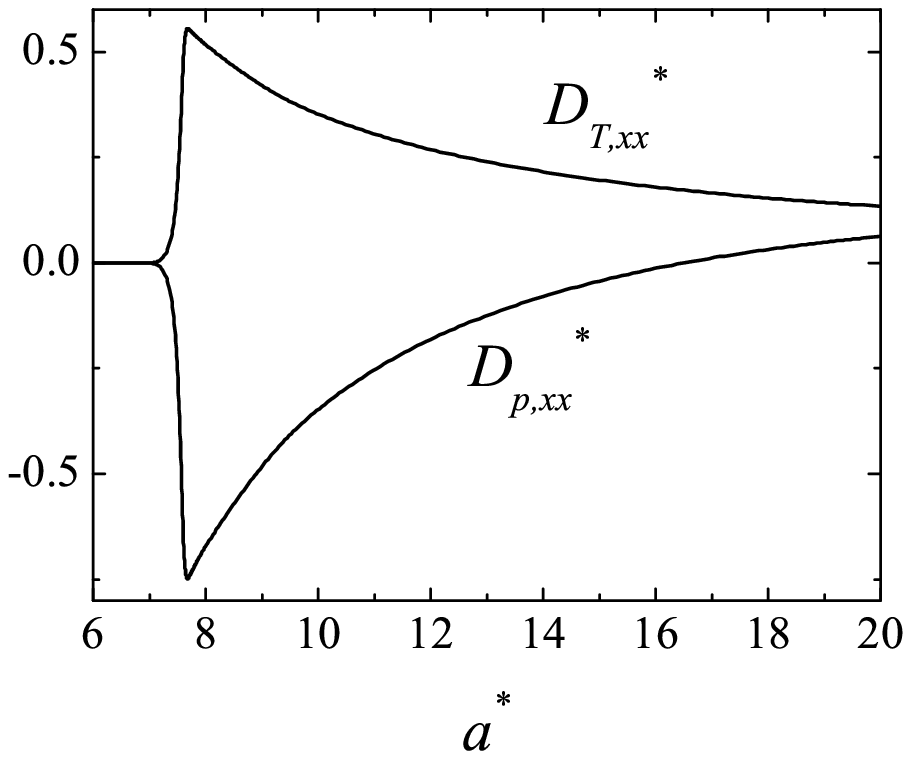}}
%&\resizebox{3.2cm}{!}
{\includegraphics[width=0.4\columnwidth]{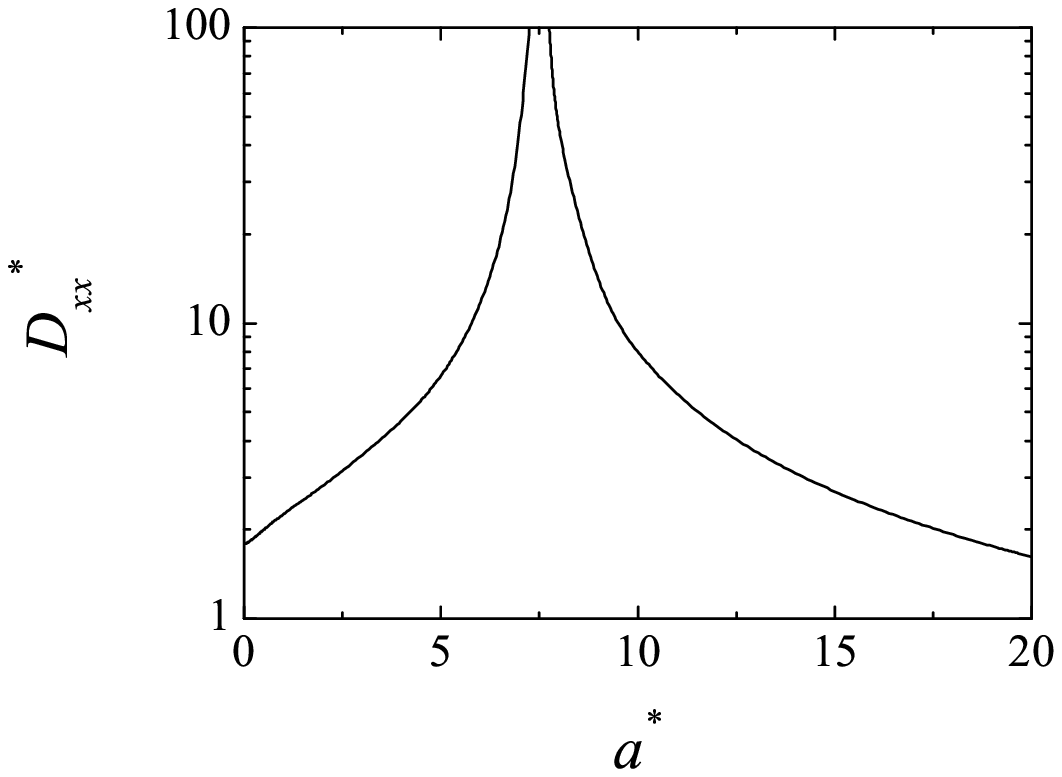}}
%\end{tabular}
%\resizebox{6.5cm}{!}{\includegraphics{fig4.eps}}
%\end{center}
\caption{Same as in Fig.\ \ref{fig1} but for the coefficients $D_{p,xx}^*$, $D_{T,xx}^*$ and $D_{xx}^*$.
\label{fig4}}
\end{figure}

%%%%%%%%%%%%%%%%%%%%%%%%%%%%%%%%%%%%%%%%%%%%%%%%%%%%%%%%%%%%%%%%%%%%%%%%%%%%%%%%%%%%%
\section{Some illustrative systems}
\label{sec5}

The results derived in the previous section gives the dependence of the set of (scaled) diffusion coefficients 
${\cal D}_{ij}\equiv \left\{D_{ij}^*, D_{p,ij}^*, D_{T,ij}^*\right\}$ in the disordered and
ordered phases in terms of the mass ratio $\mu$, the coefficients of restitution $\al_{12}$ and $\al_{22}$ and the dimensionality of the system $d$. Highly nonlinear functions on the above parameter space appear.

According to equations \eqref{5.3} and \eqref{5.5}-\eqref{5.7}, ${\cal D}_{xz}={\cal D}_{zx}={\cal D}_{yz}=
{\cal D}_{zy}=0$ in agreement with the symmetry of the linear shear flow \eqref{2.11}. Therefore,
there are five relevant (nonzero) elements of the tensors ${\cal D}_{ij}$: the three diagonal (${\cal D}_{xx}$, ${\cal D}_{yy}$ and ${\cal D}_{zz}$) and the two off-diagonal elements (${\cal D}_{xy}$
and ${\cal D}_{yx}$). Equations \eqref{5.3} and \eqref{5.5}-\eqref{5.7} also show that the anisotropy produced by the shear flow leads to the properties ${\cal D}_{xx}\neq {\cal D}_{yy}={\cal D}_{zz}$
and ${\cal D}_{xy}\neq {\cal D}_{yx}$. Note that the equality ${\cal D}_{yy}={\cal D}_{zz}$ is a consequence of the identity $P_{1,yy}^*=P_{1,zz}^*$. This property is due to the interaction model
considered since ${\cal D}_{yy}\neq {\cal D}_{zz}$ for IHS \cite{G02,G07bis}.

In order to illustrate the shear-rate dependence of those coefficients, we consider a three-dimensional system ($d=3$) with a \emph{common} coefficient of restitution ($\al\equiv \al_{22}=\al_{12}$).
This reduces our parameter space to three independent quantities: $\mu$, $\al$ and $a^*$. In this case (symmetric dissipation), according to equation \eqref{2.18.3}, the value of the threshold mass ratio
$\mu_{\text{th}}^{(-)}$ for the light impurity phase is independent of the coefficient
of restitution, i.e., $\mu_{\text{th}}= \sqrt{2}-1\simeq 0.414$. Since this phase is also present in the elastic case \cite{MSG96}, we focus our attention first onto a system with a mass ratio $\mu<\mu_{\text{th}}^{(-)}$. More specifically, we consider the mass ratio $\mu=0.2$ for which the critical value of the (reduced) shear rate
$a_c^*\simeq 7.56$ and so, the disordered phase exists for $a^*\gtrsim 7.56$.

Figures \ref{fig1}--\ref{fig4} are for the dependence of the coefficients ${\cal D}_{ij}$ for $\mu=0.2$ and $\al=0.9$. As expected, the results show that the coefficients $D_{p,ij}^*$ and $D_{T,ij}^*$ vanish in the disordered phase but they are different from zero in the ordered phase. The coefficients $D_{ij}^*$ diverge in the disordered phase at the critical point (since ${\cal P}_{1ij,\text{dis}}^{(1)}\propto \gamma_1\to \infty$), but remain finite in the ordered phase. In general, we observe that the effect of the shear flow on diffusion is quite significant, especially for the tracer diffusion coefficients $D_{ij}^*$.

We start our discussion with the diagonal terms ${\cal D}_{ii}$ ($i=x,y,z$). In the case of $D_{yy}^*$ and $D_{xx}^*$, it appears that their shear-rate dependence is qualitatively similar in the ordered phase ($a^*>a_c^*$) since both coefficients decrease with increasing the shear rate. In the disordered phase however, while $D_{yy}^*$ exhibits a non-monotonic dependence on $a^*$, $D_{xx}^*$ increases with $a^*$ and thus, shearing enhances
diffusion along the $x$ direction. In addition, at a more quantitative level, we also observe that the anisotropy of the system (as measured by the difference $D_{xx}^*-D_{yy}^*$)
grows with the shear rate in the disordered phase while the opposite happens in the ordered phase (since for instance, $D_{xx}^*-D_{yy}^* \simeq 7.29$ at $a^*=10$ and $D_{xx}^*-D_{yy}^* \simeq 4.13$
at $a^*=12$). In any case, both diagonal elements (which can be understood as generalized mutual diffusion coefficients in a sheared mixture) tend to zero as the shear rate becomes large, this tendency
being much slower in the case of $D_{xx}^*$. As far as the diagonal elements $D_{p,ii}^*$ and $D_{T,ii}^*$ are concerned, we see first that they can be positive or negative in the ordered phase, although
their magnitude is much smaller than their counterparts $D_{ii}^*$. Moreover, $|D_{p,ii}^*|$ and $|D_{T,ii}^*|$ decrease with $a^*$ and tend to vanish at large shear rates.

We consider now the off-diagonal elements ${\cal D}_{ij}$ ($i\neq j$). They measure cross effects in the diffusion of
particles induced by the shear flow. Thus, for instance, $D_{xy}^*$ gives the transport of mass along the direction of the flow ($x$ axis) due to a concentration gradient parallel to the gradient of the flow velocity ($y$ axis). While $D_{xy}^*$ and $D_{yx}^*$ are negative, a different behaviour is reported for
the coefficients $D_{p,ij}^*$ and $D_{T,ij}^*$ (with $i\neq j$). As in the case of the diagonal elements, the magnitude of the latter coefficients is in general smaller than that of the cross-coefficients $D_{xy}^*$ and $D_{yx}^*$. We also observe that the shear-rate dependence of $|D_{xy}^*|$ and $|D_{yx}^*|$ is quite similar in both phases for the system parameters chosen in figures \ref{fig2} and \ref{fig3}: they first display a non-monotonic dependence on $a^*$ in the disordered phase, then tend to infinity at the critical point while they decrease upon increasing the shear rate in the ordered phase. As for the diagonal elements, $|D_{xy}^*|$ is in general larger than $|D_{yx}^*|$ showing that the coupling between the shear field and the concentration gradient enhances significantly the mass transport along the direction of the flow. Finally, it must noted that the behaviour of $D_{p,ij}^*$ and $D_{T,ij}^*$ (with $i\neq j$) is quite similar to that of the diagonal elements since they vanish in the disordered phase and then their magnitude decreases as the shear rate increases.
\begin{figure}
%[hbtp]
%\begin{center}
%\begin{tabular}{lr}
%\resizebox{3.7cm}{!}
{\includegraphics[width=0.35\columnwidth]{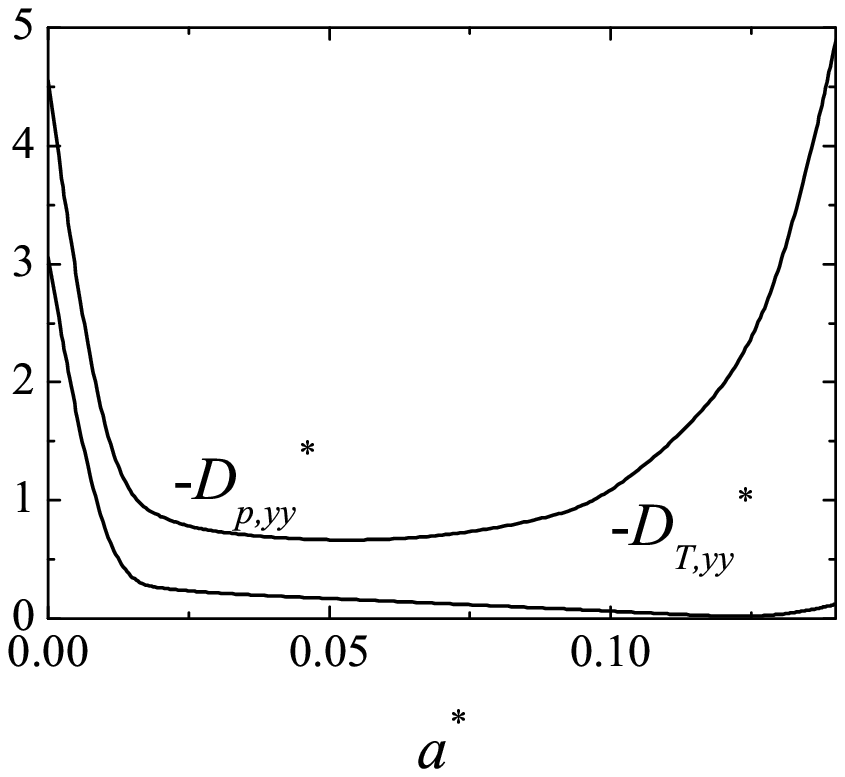}}
%&\resizebox{3.2cm}{!}
{\includegraphics[width=0.36\columnwidth]{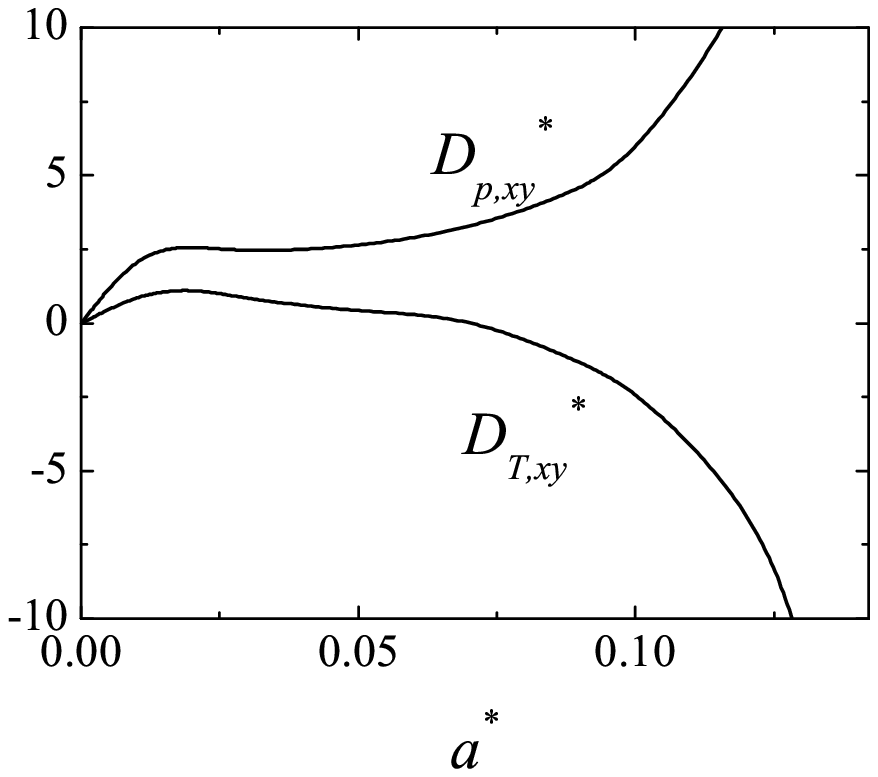}}
%\end{tabular}
%\resizebox{6.5cm}{!}{\includegraphics{fig4.eps}}
%\end{center}
\caption{Shear rate dependence of $-D_{p,yy}^*$, $-D_{T,yy}^*$, $D_{p,xy}^*$ and $D_{T,xy}^*$ for a three-dimensional system with a mass ratio $\mu=50$ and a (common) coefficient of
restitution $\al=\al_{22}=\al_{12}=0.9$. In this case, the ordered phase exists for $a^*<a^{*(+)}\simeq 0.142$. On the other hand,
in the disordered region ($a^*>0.142$), all quantities plotted vanish.
\label{fig5}}
\end{figure}
\begin{figure}
%[hbtp]
%\begin{center}
%\begin{tabular}{lr}
%\resizebox{3.7cm}{!}
{\includegraphics[width=0.35\columnwidth]{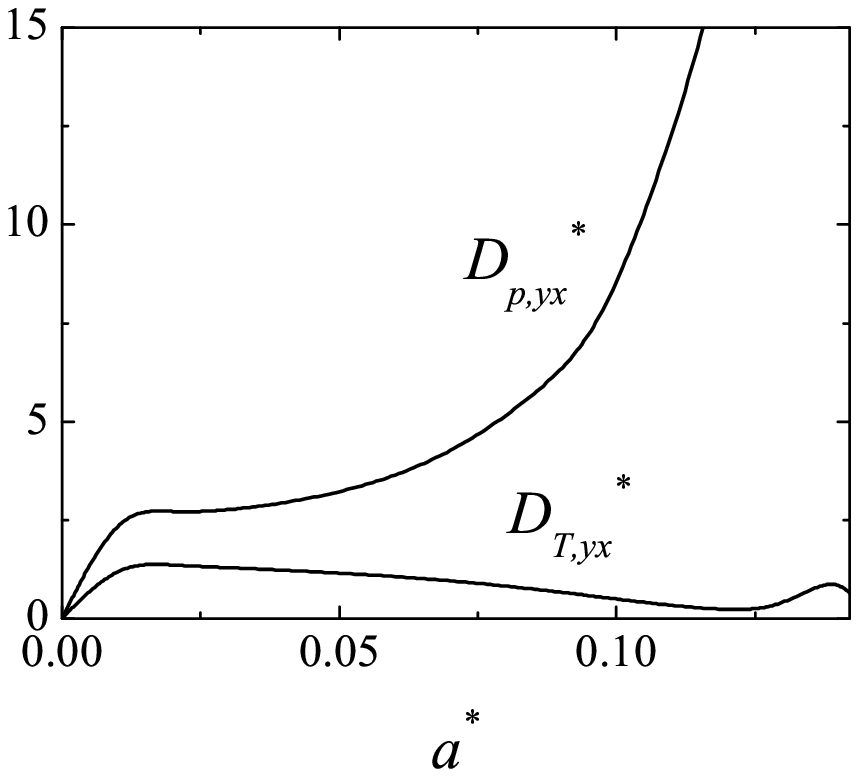}}
%&\resizebox{3.2cm}{!}
{\includegraphics[width=0.36\columnwidth]{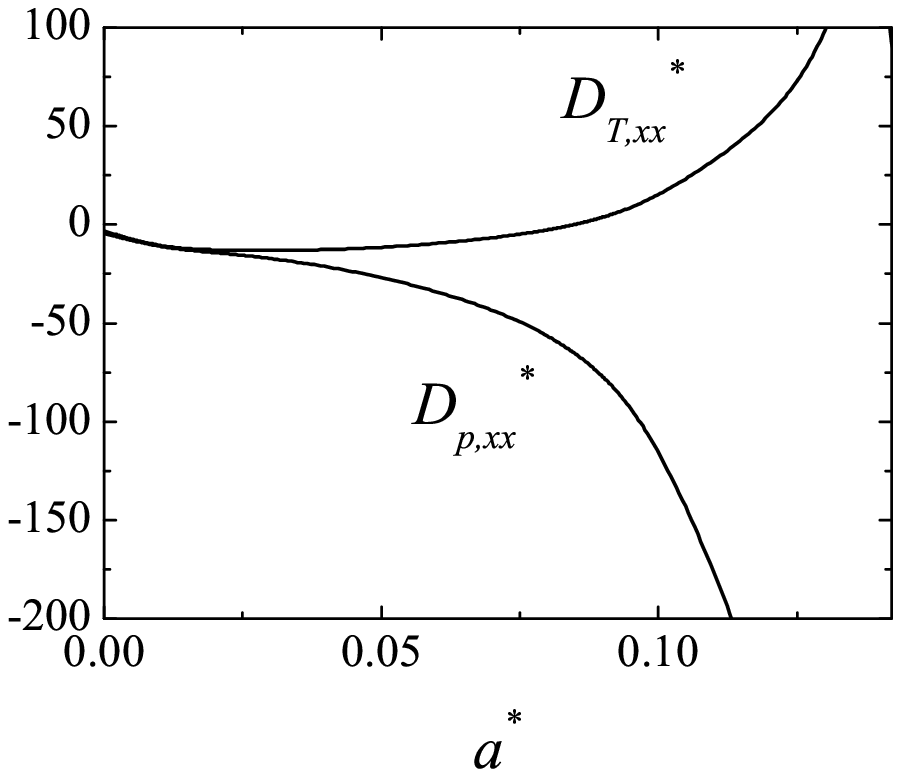}}
%\end{tabular}
%\resizebox{6.5cm}{!}{\includegraphics{fig4.eps}}
%\end{center}
\caption{Same as in Fig.\ \ref{fig5} but for the coefficients $D_{p,yx}^*$, $D_{T,yx}^*$, $D_{p,xx}^*$ and $D_{T,xx}^*$.
\label{fig6}}
\end{figure}

Now, we consider a situation where the ordered phase appears for heavy impurities.
For symmetric dissipation, this phase exists for $\mu>\mu_{\text{HCS}}^{(+)}$
[where $\mu_{\text{HCS}}^{(+)}$ is given by the second identity of equation \eqref{eq:muhcs}] and $a^*<a^{*(+)}$ [where $a^{*(+)}$ is given by equation \eqref{2.18.2}]. For $d=3$ and  $\al=\al_{22}=\al_{12}=0.9$,
$\mu_{\text{HCS}}^{(+)}\simeq 38.03$ and $a^{*(+)}\simeq 0.142$. Figures \ref{fig5} and \ref{fig6} show the shear-rate dependence of the relevant elements of the tensors $D_{p,ij}^*$ and $D_{T,ij}^*$.
The mutual diffusion tensor $D_{ij}^*$ has not been plotted since in the ordered region the conventional diffusion coefficient $D^*$ [defined by equation \eqref{4.14}] yields unphysical negative values
($D^*<0$) when one considers the ``vanilla'' version of the inelastic Maxwell model. This drawback of the model was already discussed in Ref.\ \cite{GKT15}, where it was found that the kinetic theory calculations disagree with Monte Carlo simulations: the latter predict that the coefficient $D^*$ (which can be understood as the vanishing shear rate limit of the tensor $D_{ij}^*$) diverges in the ordered phase while theoretical predictions yield finite values. It appears from figures \ref{fig5} and \ref{fig6} that the impact of shear flow on $D_{p,ij}^*$ and $D_{T,ij}^*$ is more important here than in the light impurity phase case. This could come as a surprise since the magnitude of shear rates covering the ordered heavy tracer region is smaller than that of the corresponding ordered light tracer region. However, it seems that the effect of the mass ratio on diffusion (with relatively small shear rates) in
the present case (Brownian limit) is more significant than the effect of the shear rate on diffusion (with relatively large shear rates) when the tracer particles are lighter than the gas particles.
Thus, in particular, there is a significant enhancement of the magnitude of the coefficients $D_{p,xx}^*$ and $D_{T,xx}^*$ with respect to their vanishing shear rate values $D_p^*$ [defined by
equation \eqref{4.15}] and $D_T^*$ [defined by equation \eqref{4.16}], respectively.

\section{Discussion}
\label{sec6}

In this paper, we have analyzed the effects of a recent non-equilibrium transition \cite{GT11,GT12a} found for inelastic Maxwell Models when the concentration of one of the species $x_1$ of a binary mixture is negligible (tracer limit). The emphasis was put on the diffusion coefficients of impurities immersed in a strongly sheared granular gas. In this transition, at given values of the shear rate and the parameters of the system (masses and coefficients of restitution for collisions between tracer and gas particles and gas particles among themselves), there are regions (coined as ordered phases) where quite surprisingly the relative contribution of the tracer species to the total properties of the mixture does not vanish as $x_1\to 0$. Two families of ordered phase appear: (i) a light impurity phase which exists when the mass ratio $\mu\equiv m_1/m_2$ does not exceed the threshold value $\mu_{\text{th}}^{(-)}$ [defined by equation \eqref{2.18.3}] and the shear rate is larger than a certain critical value, and (ii) a heavy impurity phase which appears when $\mu >\mu_{\text{HCS}}^{(+)}$ [defined by the second identity of equation \eqref{eq:muhcs}] and shear rates smaller than $a^{*(+)}$ [defined by equation \eqref{2.18.2}]. The light impurity phase can also exist at $a^*=0$ when $\al_{12}>\sqrt{(1+\al_{22})^2/2}$. While the light impurity phase was already found \cite{MSG96} in the case of ordinary (elastic) mixtures, the second one (heavy impurity phase) is absent for elastic collisions since $a^{*(+)}=0$ when $\al_{22}=1$. It must be noted that both light and heavy ordered phases disappear when $a^*<a_c^*$ and $\mu<\mu_{\text{HCS}}^{(+)}$ in the particular case of symmetric dissipation ($\al_{12}=\al_{22}$). As expected, in the disordered phase, the properties of the mixture coincide
with that of the excess gas.

Because of the anisotropy induced by the shear flow, tensorial quantities are required to describe mass transport. Thus, the mass flux $j_1^{(1)}$ of impurities is given by equation \eqref{3.24} where the second-rank (scaled) tensors $D_{ij}^*$, $D_{p,ij}^*$ and $D_{T,ij}^*$ obey the set of coupled algebraic equations \eqref{4.11}--\eqref{4.13} for
arbitrary concentration ($x_1\neq 0$). Starting from these general equations, the forms of those shear-rate dependent tensors have been explicitly obtained in both disordered and ordered phases,
by enforcing carefully the tracer limit. It was found that the dependence of the (scaled) diffusion coefficients on both the (reduced) shear rate $a^*$ and the parameters of the mixture (mass ratio $\mu$ and the coefficients of restitution $\al_{22}$ and $\al_{12}$) is clearly different in both phases. The pressure $D_{p,ij}^*$ and thermal $D_{T,ij}^*$ diffusion coefficients vanish in the disordered
phase while they are given by equations \eqref{5.6} and \eqref{5.7}, respectively, in the ordered phase. The expression of the mutual diffusion coefficients $D_{ij}^*$, for the disordered phase, coincides with the one derived before \cite{G03} by starting from the Boltzmann-Lorentz equation for the tracer particles. On the other hand, it is given by equation \eqref{5.8} in the ordered phase.

The results show that in general the shear-rate dependence of all the coefficients is quite complex. In particular, as happens in the Navier-Stokes description\cite{GKT15}, only the mutual diffusion
coefficients $D_{ij}^*$ diverge at the critical point. Moreover, the analysis carried out in section \ref{sec4} shows that $D_{ij}^*$ turns out negative in the ordered  heavy tracer phase for all the
range of shear rates studied. Since the diagonal elements of this tensor can be seen as a generalization of the mutual diffusion coefficient $D^*$ [defined by equation \eqref{4.14}], one could expect
that these elements should be positive. The fact that $D_{ii}^*$ is negative for $a^*<a^{*(+)}$ could be a reminiscence of the unphysical behaviour found for $D^*$ in the ordered phase for extreme values
of the mass ratio \cite{GKT15}. In addition, given that significant discrepancies were found in Ref.\ \cite{GKT15} for the tracer diffusion between theory and Monte Carlo simulations in the ordered phase,
a possible scenario to explain this disagreement could be the breakdown of hydrodynamics in the ordered phase for large mass ratios. On the other hand, beyond this region, the present results
for the set of shear-rate dependent coefficients show that all of them are well behaved and so, one could speculate that granular hydrodynamics (in the sense that all the space and time dependence of
the distribution functions occurs entirely through a functional dependence on the hydrodynamic fields) is here valid. A complete answer to this question
would require additional numerical work to measure some of these coefficients. We plan to perform Monte Carlo simulations in a sheared granular mixture by following the strategy adopted years ago by
Campbell \cite{C97} who computed the self-diffusion tensor via molecular-dynamics simulations, using particle tracking and through velocity correlations.

\acknowledgments

V. G. acknowledges support of the Spanish Government through Grant No. FIS2013-42840-P and
of the Junta de Extremadura (Spain) through Grant No. GR15104, both partially financed by FEDER funds.
V. G. and E. T. also acknowledge funding by the Investissement d'Avenir LabEx PALM program
(grant number ANR-10-LABX-0039-PALM).
%%%%%%%%%%%%%%%%%%%%%%%%%%%%%%%%%%%%%%%%%%%%%%%%%%%%%%%%%%%%%%%%%%%%%%%%%%%%%%%%%%%%%%%%%%%

\appendix
\section{Expressions of the partial pressure tensors in the USF state
\label{appA}}

In this Appendix, we display the explicit forms of the (reduced) pressure tensors $P_{r,ij}^*$ ($r=1,2$) in the USF state for arbitrary concentration ($x_1\neq 0$). First, the (global)
shear stress $P_{xy}^*$ is given by $P_{xy}^*=P_{1,xy}^*+P_{2,xy}^*$ where
\beq
\label{a1}
P_{1,xy}^*=\frac{d}{2a^*}\left[A_{12}^*-B_{12}^*-\left(B_{11}^*+\lambda-A_{11}^*+
A_{12}^*-B_{12}^*\right)p_1^*\right],
\eeq
\beq
\label{a2}
P_{2,xy}^*=\frac{d}{2a^*}\left[A_{21}^*-B_{21}^*-\left(B_{22}^*+\lambda-A_{22}^*+
A_{21}^*-B_{21}^*\right) (1-p_1^*)\right].
\eeq
Here,
\begin{equation}
\label{a3}
A_{11}^*=\frac{\omega_{11}^*}{2(d+2)}(1+\alpha_{11})^2+\frac{\omega_{12}^*}{d+2}
\mu_{21}^2(1+\alpha_{12})^2,
\end{equation}
\begin{equation}
\label{a4}
A_{12}^*=\frac{\omega_{12}^*}{d+2}\frac{\rho_1}{\rho_2}\mu_{21}^2(1+\alpha_{12})^2,
\end{equation}
\beq
\label{a5}
B_{11}^*=\frac{\omega_{11}^*}{d(d+2)}(1+\alpha_{11})(d+1-\alpha_{11})
+\frac{2\omega_{12}^*}{d(d+2)}\mu_{21}(1+\alpha_{12})
\left[d+2-\mu_{21}(1+\alpha_{12})\right],
\eeq
\begin{equation}
\label{a6}
B_{12}^*=-\frac{2}{d}A_{12}^*,
\end{equation}
where $\omega_{rs}^*=\omega_{rs}^*/\nu_0$. Adequate change of indices ($1\leftrightarrow 2$) provide the equations pertaining to $A_{22}^*$, $A_{21}^*$, $B_{22}^*$, and $B_{21}^*$. In addition,
in equations \eqref{a1} and \eqref{a2} the energy ratio $p_1^*$ can be written as \cite{GT12a}
\begin{equation}
\label{c0}
p_1^*=\frac{K a^{*2}+L}{R a^{*2}+S},
\end{equation}
where
\begin{widetext}
\begin{equation}
\label{c0.1}
K=-2A_{12}^*\lambda^2+4(A_{22}^*B_{12}^*-A_{12}^*B_{22}^*)\lambda+2A_{22}^*
B_{12}^*(B_{11}^*+B_{22}^*)-2A_{12}^*(B_{12}^*B_{21}^*+B_{22}^{*2}),
\end{equation}
\begin{equation}
\label{c0.2}
L=d(B_{12}^*-A_{12}^*)\left[\lambda^2+(B_{11}^*+B_{22}^*)\lambda+B_{11}^*B_{22}^*-B_{12}^*B_{21}^*\right]^2,
\end{equation}
\beqa
\label{c0.3}
R&=&2(A_{11}^*-A_{12}^*)\lambda^2-4\left[B_{12}^*(A_{21}^*-A_{22}^*)+B_{22}^*(A_{12}^*
-A_{11}^*)\right]\lambda \nonumber\\
& & +2B_{12}^*(B_{11}^*+B_{22}^*)(A_{22}^*-A_{21}^*)+2(A_{11}^*-A_{12}^*)
(B_{12}^*B_{21}^*+B_{22}^{*2}),
\eeqa
\begin{equation}
\label{c0.4}
S=d(A_{11}^*-A_{12}^*-B_{11}^*+B_{12}^*-\lambda)\left[\lambda^2+(B_{11}^*+B_{22}^*)
\lambda+B_{11}^*B_{22}^*-B_{12}^*B_{21}^*\right]^2.
\end{equation}
\end{widetext}

The other relevant element of the pressure tensor is $P_{yy}^*=P_{zz}^*$. It is given by $P_{yy}^*=P_{1,yy}^*+P_{2,yy}^*$ where
\begin{widetext}
\begin{equation}
\label{a7}
P_{1,yy}^*=P_{1,zz}^*=\frac{(B_{22}^*+\lambda)\left[p_1^*A_{11}^*+(1-p_1^*)A_{12}^*
\right]
-B_{12}^*\left[p_1^*A_{21}^*+(1-p_1^*)A_{22}^*\right]}{(B_{11}^*+\lambda)
(B_{22}^*+\lambda)-
B_{12}^*B_{21}^*},
\end{equation}
\begin{equation}
\label{a8}
P_{2,yy}^*=P_{2,zz}^*=\frac{(B_{11}^*+\lambda)\left[p_1^*A_{21}^*+(1-p_1^*)
A_{22}^*\right]
-B_{21}^*\left[p_1^*A_{11}^*+(1-p_1^*)A_{12}^*\right]}{(B_{11}^*+\lambda)
(B_{22}^*+\lambda)-
B_{12}^*B_{21}^*}.
\end{equation}
\end{widetext}
Finally, the $xx$-element $P_{xx}^*=P_{1,xx}^*+P_{2,xx}^*$ where its partial contributions can be easily determined from the constraint
\beq
\label{p1xx}
P_{r,xx}^*=d x_r \gamma_r-(d-1)P_{r,yy}^*,
\eeq
where $\gamma_r\equiv T_r/T$ is the partial temperature of species $r$.

The above expressions for the partial contributions $P_{r,ij}^*$ to the pressure tensor hold for arbitrary values of $x_1$.  Let us consider now the forms of $P_{r,ij}^*$ in the tracer
limit ($x_1\to 0$). In this case, we assume that $P_{r,ij}^*$ and $p_1^*$ can be expanded as
\begin{equation}
\label{a9}
P_{r,ij}^*={\cal P}_{r,ij}^{(0)}+{\cal P}_{r,ij}^{(1)}x_1
+{\cal P}_{r,ij}^{(2)}x_1^2+\ldots,
\end{equation}
\begin{equation}
\label{a10}
p_{1}^*=p_{1}^{(0)}+p_{1}^{(1)}x_1+p_{1}^{(2)}x_1^2+\ldots.
\end{equation}
The expressions of ${\cal P}_{r,ij}^{(k)}$ and $p_1^{(k)}$ will be different if $\lambda_2^{(0)}>\lambda_1^{(0)}$ (disordered phase) or $\lambda_1^{(0)}>\lambda_2^{(0)}$ (ordered phase).
In particular, in the lowest order in $x_1$, the expressions of ${\cal P}_{r,ij}^{(0)}$ in the disordered phase are simply ${\cal P}_{1xy,\text{dis}}^{(0)}={\cal P}_{1yy,\text{dis}}^{(0)}=0$,
\begin{equation}
\label{a11}
{\cal P}_{2xy,\text{dis}}^{(0)}=-\frac{A_{22}^{(0)}}{(B_{22}^{(0)}
+\lambda_2^{(0)})^2}a^*,
\end{equation}
\begin{equation}
\label{a12}
{\cal P}_{2yy,\text{dis}}^{(0)}=\frac{A_{22}^{(0)}}{B_{22}^{(0)}+\lambda_2^{(0)}},
\end{equation}
where
\begin{equation}
\label{a13}
A_{22}^{(0)}=\frac{(1+\alpha_{22})^2}{2(d+2)},
\end{equation}
\begin{equation}
\label{a14}
B_{22}^{(0)}=\frac{(1+\alpha_{22})(d+1-\alpha_{22})}{d(d+2)},
\end{equation}
and use has been made of the fact that $p_1^{(0)}=0$ in the disordered phase. In addition, to get the (reduced) diffusion tensor $D_{ij}^*$ in the disordered phase, we need the expressions of
${\cal P}_{1xy,\text{dis}}^{(1)}$ and ${\cal P}_{1yy,\text{dis}}^{(1)}$. They are given by
\begin{equation}
\label{a14.1}
{\cal P}_{1xy,\text{dis}}^{(1)}=\frac{d}{2a^*}\left[A_{12}^{(1)}-
B_{12}^{(1)}+\left(A_{11}^{(0)}-
B_{11}^{(0)}-\lambda_2^{(0)}\right)\gamma_1\right],
\end{equation}
\begin{equation}
\label{a14.2}
{\cal P}_{1yy,\text{dis}}^{(1)}=\frac{(B_{22}^{(0)}+
\lambda_2^{(0)})\left(A_{11}^{(0)}\gamma_1
+A_{12}^{(1)}\right)-A_{22}^{(0)}B_{12}^{(1)}}{(B_{11}^{(0)}+\lambda_2^{(0)})
(B_{22}^{(0)}+\lambda_2^{(0)})},
\end{equation}
where
\beq
\label{a110}
A_{11}^{(0)}=\frac{\mu_{21}^2}{d+2}(1+\alpha_{12})^2,
\eeq
\beq
\label{b110}
B_{11}^{(0)}=\frac{2}{d(d+2)}\mu_{21}(1+\alpha_{12})
\left[d+2-\mu_{21}(1+\alpha_{12})\right],
\eeq
\beq
\label{a121}
A_{12}^{(1)}=\frac{\mu_{12}\mu_{21}}{(d+2)}(1+\alpha_{12})^2,
\eeq
\vspace{0.05mm}
\beq
\label{b121}
B_{12}^{(1)}=-\frac{2}{d}A_{12}^{(1)}.
\eeq
On the other hand, the temperature ratio $\gamma_1=T_1/T$ is, in the tracer limit,
\begin{equation}
\label{a14.3}
\gamma_1=\frac{D_1(\lambda_2^{(0)})}{\Delta_0(\lambda_2^{(0)})},
\end{equation}
where the functions $D_1(\lambda)$ and $\Delta_0(\lambda)$ are given by
\begin{widetext}
\beq
\label{D1}
D_1(\lambda)=d(B_{12}^{(1)}-A_{12}^{(1)})(B_{11}^{(0)}+\lambda)^2(B_{22}^{(0)}+\lambda)^2
+2a^{*2}\left[
A_{22}^{0)}B_{12}^{(1)}(B_{11}^{(0)}+B_{22}^{(0)}+2\lambda)-A_{12}^{(1)}(B_{22}^{(0)}
+\lambda)^2\right],
\eeq
\beq
\label{delta0}
\Delta_0(\lambda)=(B_{22}^{(0)}+\lambda)^2\left[ 2a^{*2}A_{11}^{(0)}+d(A_{11}^{(0)}-B_{11}^{(0)}-
\lambda)(B_{11}^{(0)}+\lambda)^2\right].
\eeq
\end{widetext}
The quantities ${\cal P}_{1xx,\text{dis}}^{(0)}$ and ${\cal P}_{1xx,\text{dis}}^{(1)}$ can be easily identified from the relation \eqref{p1xx} with the result
\beq
\label{p1xxdis0}
{\cal P}_{1xx,\text{dis}}^{(0)}=-(d-1){\cal P}_{1yy,\text{dis}}^{(0)},
\eeq
\beq
\label{p1xxdis1}
{\cal P}_{1xx,\text{dis}}^{(1)}=d\gamma_1-(d-1){\cal P}_{1yy,\text{dis}}^{(1)}.
\eeq

In the ordered phase, the zeroth-order expressions  for the elements of the tracer pressure tensor $P_{1,ij}^*$ are
\begin{equation}
\label{a15}
{\cal P}_{1xy,\text{ord}}^{(0)}=\frac{d}{2a^*}\left(A_{11}^{(0)}-
B_{11}^{(0)}-\lambda_1^{(0)}\right)p_1^{(0)},
\end{equation}
\begin{equation}
\label{a16.1}
{\cal P}_{1yy,\text{ord}}^{(0)}=\frac{A_{11}^{(0)}}
{B_{11}^{(0)}+\lambda_1^{(0)}}p_1^{(0)},
\end{equation}
where
\begin{equation}
\label{a17}
A_{21}^{(0)}=A_{12}^{(1)}=\frac{\mu_{12}\mu_{21}}{(d+2)}(1+\alpha_{12})^2,
\quad B_{21}^{(0)}=B_{12}^{(1)}=-\frac{2}{d}A_{21}^{(0)},
\end{equation}
\begin{equation}
\label{a18}
A_{11}^{(1)}=\frac{1}{2(d+2)}(1+\alpha_{11})^2-A_{11}^{(0)},
\end{equation}
\begin{equation}
\label{a19}
B_{11}^{(1)}=\frac{1}{d(d+2)}(1+\alpha_{11}) (d+1-\alpha_{11})-B_{11}^{(0)},
\end{equation}
\begin{equation}
\label{a20}
B_{22}^{(1)}=\frac{2}{d(d+2)}\mu_{12}(1+\alpha_{12})\left[d+2-\mu_{12}
(1+\alpha_{12})\right]
-B_{22}^{(0)},
\end{equation}
\begin{equation}
\label{a21}
A_{22}^{(1)}=\frac{\mu_{21}^2}{d+2}(1+\alpha_{12})^2-A_{22}^{(0)}.
\end{equation}
The zeroth-order contribution $p_1^{(0)}$ to the energy ratio is \cite{GT12a}
\begin{equation}
\label{c7}
p_1^{(0)}=\frac{D_1(\lambda_1^{(0)})}{\Delta_0'(\lambda_1^{(0)})\lambda_1^{(1)}+
\Delta_1(\lambda_1^{(0)})},
\end{equation}
where
\beq
\label{c7.1}
\Delta_0'(\lambda_1^{(0)})\equiv\left(\frac{\partial \Delta_0(\lambda)}{\partial
\lambda}\right)_{\lambda=\lambda_1^{(0)}},
\eeq
and the expression of $\lambda_1^{(1)}$ is given by equations (C8)--(C13) of Ref.\ \cite{GT12a}.

The first-order corrections ${\cal P}_{1xy,\text{ord}}^{(1)}$ and ${\cal P}_{1yy,\text{ord}}^{(1)}$ are also needed to determine the diffusion coefficients $D_{ij}^*$ in the ordered phase.
Their evaluation is quite involved and can be obtained by expanding the expressions \eqref{a1} and \eqref{a7} up to first order in $x_1$. After some algebra, one gets
\begin{widetext}
\begin{equation}
\label{a23}
{\cal P}_{1xy,\text{ord}}^{(1)}=\frac{d}{2a^*}\left[A_{12}^{(1)}-B_{12}^{(1)}-
\left(B_{11}^{(1)}+\lambda_1^{(1)}-A_{11}^{(1)}+A_{12}^{(1)}-B_{12}^{(1)}\right)p_1^{(0)}
-\left(B_{11}^{(0)}+\lambda_1^{(0)}-A_{11}^{(0)}\right)p_1^{(1)}\right],
\end{equation}
\begin{eqnarray}
\label{a24}
{\cal P}_{1yy,\text{ord}}^{(1)}&=&\frac{1}{(B_{11}^{(0)}+\lambda_1^{(0)})^2
(B_{22}^{(0)}+\lambda_1^{(0)})}\left\{\left(A_{12}^{(1)}(1-p_1^{(0)})+A_{11}^{(1)}p_1^{(0)}+A_{11}^{(0)}p_1^{(1)}\right)
(B_{11}^{(0)}+\lambda_1^{(0)})(B_{22}^{(0)}+\lambda_1^{(0)})\right.\nonumber\\
&+&(B_{11}^{(0)}+\lambda_1^{(0)})\left[A_{11}^{(0)}p_1^{(0)}\left(B_{22}^{(1)}+\lambda_1^{(1)}\right)-B_{12}^{(1)}\left(
A_{22}^{(0)}(1-p_1^{(0)})+A_{21}^{(0)}p_1^{(0)}\right)\right]\nonumber\\
&-&\left. A_{11}^{(0)}p_1^{(0)}\left[(B_{11}^{(1)}+\lambda_1^{(1)})(B_{22}^{(0)}+\lambda_1^{(0)})+
(B_{11}^{(0)}+\lambda_1^{(0)})(B_{22}^{(1)}+\lambda_1^{(1)})-B_{21}^{(0)}B_{12}^{(1)}
\right]\right\}.
\end{eqnarray}
\end{widetext}
The explicit form of the first-order correction $p_1^{(1)}$ to the energy ratio is displayed in Appendix \ref{appC}. Finally, the quantities ${\cal P}_{1xx,\text{ord}}^{(0)}$ and
${\cal P}_{1xx,\text{ord}}^{(1)}$ are defined as
\beq
\label{p1xxord0}
{\cal P}_{1xx,\text{ord}}^{(0)}=d p_1^{(0)}-(d-1){\cal P}_{1yy,\text{ord}}^{(0)},
\eeq
\beq
\label{p1xxord1}
{\cal P}_{1xx,\text{ord}}^{(1)}=d p_1^{(1)}-(d-1){\cal P}_{1yy,\text{ord}}^{(1)}.
\eeq

%%%%%%%%%%%%%%%%%%%%%%%%%%%%%%%%%%%%%%%%%%%%%%%%%%%%%%%%%%%%%%%%%%%%%%%%%%%%%

\section{First-order correction to the energy ratio}
\label{appC}

In the general case ($x_1 \neq 0$), the energy ratio $p_1^*$ is given by equation \eqref{c0}. In the tracer limit ($x_1\to 0$), $p_1^*$ can be expanded in powers of $x_1$ as in equation \eqref{a10}
and $p_1^{(0)}$ is defined by equation \eqref{c7}. In this Appendix, we want to get its first-order correction $p_1^{(1)}$. If $x_1\to 0$, the energy ratio $p_1^*$ becomes
\begin{equation}
\label{c1}
p_1^*(\lambda,a^*)\approx x_1 \frac{D_1(\lambda,a^*)+D_2(\lambda,a^*)x_1}{\Delta_0(\lambda,a^*)+\Delta_1(\lambda,a^*)x_1+
\Delta_2(\lambda,a^*)x_1^2},
\end{equation}
where the dependence on $\mu$, $\alpha_{11}$, $\alpha_{22}$, and $\alpha_{12}$ is implicitly assumed on the right-hand side of equation \eqref{c1}. The expressions of $D_1$ and $\Delta_0$ are given by
equations \eqref{D1} and \eqref{delta0}, respectively, while  $D_2$, $\Delta_1$ and $\Delta_2$ can be easily obtained from the general form of $p_1^*$. Their explicit expressions are too cumbersome
to be provided here and will be omitted \cite{code}. Equation \eqref{c1} applies for $\lambda_1$ and $\lambda_2$. The expansion of $\lambda_1$ in powers of $x_1$ can be written as
\begin{equation}
\label{c2} 
\lambda_1(a^*,x_1)\approx \lambda_1^{(0)}(a^*)+\lambda_1^{(1)}(a^*) x_1+\lambda_1^{(2)}(a^*) x_1^2,
\end{equation}
where the quantities $\lambda_1^{(1)}(a^*)$ and $\lambda_1^{(2)}(a^*)$ can be obtained from the sixth-degree polynomial equation defining $\lambda$ in the USF problem \cite{GT12a}. As said before,
the expression of $\lambda_1^{(1)}$ is given by equations (C8)--(C13) of Ref.\ \cite{GT12a}. On the other hand, the expression of $\lambda_1^{(2)}(a^*)$ is also too large to be displayed here.

The form of $p_1^{(1)}$ can be obtained by taking the tracer limit in equation \eqref{c1}. After some algebra, one arrives at
\begin{widetext}
\beq
\label{p11}
p_1^{(1)}=\frac{p_1^{(0)}}{D_1(\lambda_1^{(0)})}\left[D_2(\lambda_1^{(0)})+
D_1'(\lambda_1^{(0)})\lambda_1^{(1)}
- p_1^{(0)}\left(\Delta_2 (\lambda_1^{(0)})+\Delta_0'(\lambda_1^{(0)})\lambda_1^{(2)}
+\Delta_1'(\lambda_1^{(0)})\lambda_1^{(1)}+
\frac{1}{2}\Delta_0''(\lambda_1^{(0)})
\lambda_1^{(1)2}\right)\right],
\eeq
where
\begin{equation}
\label{c5}
D_1'(\lambda_1^{(0)})\equiv\left(\frac{\partial D_1(\lambda)}{\partial
\lambda}\right)_{\lambda=\lambda_1^{(0)}}, \quad \Delta_1'(\lambda_1^{(0)})\equiv\left(\frac{\partial \Delta_1(\lambda)}{\partial
\lambda}\right)_{\lambda=\lambda_1^{(0)}}, \quad \Delta_0''(\lambda_1^{(0)})\equiv\left(\frac{\partial^2 \Delta_0(\lambda)}{\partial
\lambda^2}\right)_{\lambda=\lambda_1^{(0)}}.
\eeq
\end{widetext}

In the absence of shear ($a^*=0$), the expression \eqref{p11} for $p_1^{(1)}$ reduces to the one previously derived in the homogeneous cooling state \cite{GKT15}. This shows the consistency of our results.

%%%%%%%%%%%%%%%%%%%%%%%%%%%%%%%%%%%%%%%%%%%%%%%%%%%%%%%%%%%%%%%%%%%%%%%%%%%%%%%%%%%%%%%%%%%%%%%%%%%


\begin{thebibliography} {99}
\bibitem{note}While we were working on this paper, we have realized that there were some typos in the expressions given in Ref.\ \cite{GT12a} for the coefficients $A_{11}^{(1)}$, $A_{22}^{(1)}$ and $B_{22}^{(1)}$. The forms displayed here for these quantities correct the above typos. On the other hand, the final expression of the energy ratio $E_1/E$ provided in Ref.\ \cite{GT12a} is correct.
\end{thebibliography}
\end{document}